\documentstyle[emulateapj,epsf]{article}

\def\g{\gamma}
\def\gb{{\mbox{\boldmath$ \gamma$}}}
\def\ob{{\mbox{\boldmath$ \omega$}}}

\def\be{\begin{equation}}
\def\ee{\end{equation}}

\def\ltorder{\hbox{ \rlap{\raise 0.425ex\hbox{$<$}}\lower
0.65ex\hbox{$\sim$} }} 
\def\gtorder{\hbox{ \rlap{\raise 0.425ex\hbox{$>$}}\lower
0.65ex\hbox{$\sim$} }} 
\begin{document}
\lefthead{Perturbations of spherical stellar systems}
\righthead{E.Vesperini, Martin D. Weinberg}
\title{Perturbations of spherical stellar systems during fly-by
encounters}

\author{E. Vesperini, Martin D. Weinberg}
\affil{Department of Physics and Astronomy, University of Massachusetts,
Amherst, MA 01003-4525, USA}
\begin{abstract}
We study the internal response of a galaxy to an unbound 
encounter and present a survey of orbital parameters  covering typical
encounters in different galactic environments. Overall, we conclude
that relatively weak encounters by low-mass interloping galaxies can
cause observable distortions in the primaries. The resulting
asymmetries may persist long after the interloper is evident.
  
We focus our attention on the production of structure in  dark halos and in
cluster ellipticals. Any distortion produced in a dark
halo can distort the embedded stellar disk, possibly leading
to the formation of lopsided and warped disks. We show that distant 
encounters with pericenters in the outer regions of a halo can 
excite strong and persistent features in the 
inner regions.
Features excited in an elliptical are
directly observable and we  predict that asymmetries
in the morphologies of these systems can be produced by relatively
small perturbers. For example, a fly-by on an orbit with pericenter equal to
the half-mass radius of the primary system and velocity of 200 km/s
(a value typical for groups) can result in a significant dipole
distortion for perturbers with  mass as small as 5\% of 
primary's mass.

We use these detailed results to predict the distribution of the $A$ parameter
defined by Abraham et al. (sensitive to lopsidedness) and the
shift between the center of mass of the primary system and the
position of the peak of density for a range of environments. 
We find that high-density, low-velocity dispersion
environments are more likely to host galaxies with significant
asymmetries. Our distribution for the $A$
parameter is in good agreement with the range spanned by the observed
values for local galaxy clusters and for distant galaxies in the Medium Deep
Survey and in the Hubble Deep Field. Assuming that primordial galaxies
were located in dense environments with previrialized low velocity
dispersions, our conclusions are consistent
with the observational results showing a systematic
trend for galaxies at larger redshifts to be more asymmetric than
local galaxies. 

Finally, 
we propose a generalized asymmetry  parameter $A(r)$ which
provides detailed information on the radial structure of the
asymmetry produced by the mechanism explored in our work.
\end{abstract}

\section{Introduction}
Galaxy interactions are likely to play a key role in determining the
morphology and the structural properties of galaxies and in driving
 evolution at all epochs and in widely-varying environments. The importance of
environmental conditions on the properties of populations of galaxies
has been recognized for a long time (Hubble \& Humason 1931) and 
observational studies have now firmly established  
relationships between the relative abundance of galaxies of different
morphological 
types and the density of the environment (see e.g. Dressler 1980,
Dressler et al. 1997), their location
in clusters (Oemler 1974) as well as the evolution of the
morphology of galaxies in clusters at different epochs 
(Butcher \& Oemler 1978, 1984).
In particular, the ratio of early to late-type galaxies has been shown to
be an increasing function of density  and a decreasing function of the
distance from the center of a cluster. 
In addition, HST
observations show  that the relative  number of
spirals to elliptical and S0 galaxies tend to increase in distant
clusters, pointing to a time evolution of the morphology of galaxies
from spirals to ellipticals (see e.g. Dressler et al. 1994a, 1994b, Couch et
al. 1998). Altogether, these results suggest that
environmental disturbances due to interactions with other
galaxies and with a cluster tidal field can lead to dramatic
changes in the structure of galaxies.

Simulations have explored strong encounters and mergers. Our recent
work suggests that relatively weak encounters can give rise to
significant asymmetries as well.
To study weaker galaxy interactions and quantify their importance,
we compute the response of a spherical stellar system
to the perturbation induced by another system 
during a fly-by, focusing our attention on the
morphology and the persistence of the distortions produced
during the encounter. The range of orbital parameters considered
encompasses the typical environmental conditions for field galaxies,
for galaxies in groups and in clusters. Our perturbative solution for these distortions
provides quantitative and qualitative predictions in low to moderate
amplitude that are difficult to reach by simulation. This approach
nicely reveals the underlying dynamics without being masked or swamped
by n-body fluctuation noise. On the other hand, numerical feasibility
restricts the complexity of possible geometries and the perturbative
assumption limits its validity for large amplitudes.

Our investigation is relevant for galactic dark halos or for 
spheroidal galaxies over a range of environments.
We explore the possibility that the halo distortions produced during a
fly-by could in turn lead to a
significant and observable distortion in an embedded stellar disk
(see e.g. Weinberg  1995, 1998b, Murali \& Tremaine 1998, Murali 1998).   
In particular the deformations produced in a dark halo  during an
encounter could be one of the culprits for distorted 
morphologies like those observed in lopsided and warped galactic disks
(see e.g.  Richter \& Sancisi 1994, Zaritsky \&
Rix 1997, Rudnick \& Rix 1998, Swaters et al. 1998, Haynes et
al. 1998, Kornreich, Haynes \& Lovelace 1998). As recently shown  by 
Weinberg (1995, 1998b),  this could be the case for the Milky Way in which
the deformations induced in the dark halo by the Magellanic
Clouds could be responsible for the observed warp in the Galactic disk.
In a recent investigation, Reshetnikov \& Combes (1998, 1999) have
found that about 40\% of a sample of 540 galaxies have warped planes
and their analysis revealed that warped galaxies tend to be located in
denser environments supporting a scenario in which interactions would
play an important role in the formation of warps.

Distortions of spheroidal galaxies are directly observable.
Our results show that in
low velocity dispersion and dense environments, where low-velocity close
encounters are more likely to occur,  
even low-mass perturbers can lead to significant asymmetries in
the primary galaxy.  Recent morphological studies of both local and
very distant 
galaxies have yielded comparisons  
of the frequencies  of
systems with distorted morphologies  in
different environments and at different distances (see e.g. Zepf \&
Whitmore 1993, Mendes de Oliveira \& Hickson 1994, Abraham et al.
1996a,1996b, van den Bergh et al. 1996, Naim, Ratnatunga \& Griffiths
1997, Brinchmann et al. 1998, Marleau \& Simard 1998, Conselice \&
Bershady 1998, Conselice \& Gallagher 1999).
On the theoretical side, the efforts have been largely concentrated on
 numerical $N$-body simulations aimed at investigating the fate and
the evolution of merging galaxies (see e.g. Barnes \& Hernquist 1992,
Barnes 1998 and references therein) and the distortions
and the morphological alterations produced in galaxies  in clusters
and groups  by the ensemble of ``tidal shocks'' with other members
(Moore et al. 1996, 1998). This mechanism can drive 
the evolution of spiral galaxies into spheroidal systems and it is
a strong candidate explanation for the difference in the
relative number of spiral to elliptical galaxies in clusters at
different redshifts.

The plan of the paper is the following.
In \S 2 we outline the method adopted for this investigation while the
mathematical details are described in the Appendix. In
\S 3 we show the results obtained for initial conditions
typical of galactic dark halos and their dependence on the orbital
parameters of the perturber. In \S 4 initial conditions relevant
for spheroidal galaxies and a survey of orbital parameters of the
perturber typical of different environments are considered. 
Overall, our results quantify the relationship between the interaction
rate for the 
environment and the probability of
observing irregular morphologies, aiding interpretation of any
trend in the fraction of asymmetric galaxies at
different redshifts. This latter aspect is of particular interest as
data on the morphology of distant galaxies are now available
from HST observations in the Hubble Deep Field and in the Medium
Deep Survey projects (Abraham et al. 1996a, 1996b). 
In order to facilitate
the comparison of our theoretical results with observational data, we
will quantify the distortions produced by means of the asymmetry
parameter $A$ (Abraham et al. 1996a) which has been determined
for a number of local and distant galaxies. We will also introduce a
generalized asymmetry parameter $A(r)$ which will provide an important
information on the radial structure of the asymmetry produced by the
mechanism we have considered and which will allow to test the theory
presented here.
The main
conclusions are summarized and discussed in \S 5.  
\section{Method}
This section outlines the perturbative solution of the
collisionless Boltzmann equation used to predict the dynamical
response of galaxies to interactions. Additional 
mathematical detail is presented in the Appendix.

The linear
perturbation theory used here has been applied in the past both to
investigate the 
collisionless stability of stellar disks (Kalnajs 1977) and of spherical
stellar systems 
(Polyachenko \& Shukhman 1981, Palmer \& Papaloizou 1987, Bertin \&
Pegoraro 1989, Saha 1991,
Weinberg 1991, Bertin et al. 1994; see also Palmer 1994 and references
therein) and to study the response of a
galaxy to perturbations 
induced by satellite companions (Weinberg 1989, 1995, 1998b).
This method is numerically intensive  but it avoids the problems of
n-body simulations due to the finite 
number of particles which can introduce spurious noise effects (see
Weinberg 1998a) and it guarantees the resolution of resonances that 
lead to the excitation of
patterns in the primary system. With too few particles, for example,
noise can cause orbital energies and angular momenta to drift on a
time scale shorter than the pattern speed of the global response of
interest. This eliminates the possibility of observing this structure
in the simulation. No doubt, real galaxies are not smooth and inherent
fluctuations may be important to their evolution. However, many
millions of n-body particles are required to approach the estimates of
natural amplitudes (Nelson \& Tremaine 1996, Weinberg 1998a).

We compute the perturbation induced on a spherical stellar system
by a point mass perturber \footnote{Although this approach can treat a
perturber with any density profile, there is no change in the global
response if the interloper is replaced by a point mass.} with
mass $m$ on a rectilinear trajectory with pericenter $p$. The response
of the primary system will be determined by the simultaneous solution
of the linearized collisionless Boltzmann and Poisson's
equations,
\be
{\partial f_1\over \partial t} +{\partial f_1\over \partial {\bf w}}
{\partial H_0\over \partial {\bf I}}- {\partial f_0\over \partial
{\bf I}} {\partial H_1\over \partial {\bf w}}=0 \label{veq},
\ee
\be
\nabla^2 \Phi_1=4\pi G\rho_1,
\ee
where the subscript $0$ denotes the equilibrium quantities, subscript
$1$ denotes the
first order perturbation of the distribution function $f$,  the 
Hamiltonian $H$, and  the potential $\Phi$. The Boltzmann equation
has been written in terms of action-angle
variables $\bf {I}, \bf{w}$. 
The perturbed potential, $\Phi_1$, is the sum of the 
tidal potential due to the perturber, $\Phi_{p}$, and the gravitational
potential of the response  to the perturbation, $\Phi_{resp}$, and
$\rho_1=\rho_1^{p}+\rho_1^{resp}$. 
The coupled Boltzmann-Poisson equations lead to an
integrodifferential equation for the potential. This can be solved
by performing a Fourier transform in the angle variables, a Laplace
transform in the time variable and expanding the perturbed quantities
in terms of spherical harmonics and of a bi-orthogonal
density-potential basis functions for the angular and the radial parts
respectively. 
The  integrodifferential
equation is then reduced to an algebraic equation for the vector of
coefficients of the expansion of the response potential and density, {\bf a},
\be
{\bf a}={\bf R}({\bf a}+{\bf b}) \label{mateq}.
\ee
The matrix {\bf R} contains the time dependence of the perturbation and the
equilibrium properties of the perturbed system and
{\bf b} is the vector of the coefficients of the expansion of the external
perturbing potential. 

It is clear from equation (\ref{mateq}) that the
self-consistent response of the perturbed system is determined both by
the perturbation applied by the external perturber and by the reaction
of the system  to its own response to this perturbation.
In general, 
an additional contribution to the total
response comes from excitation of the discrete damped
modes of the primary system besides the effects of the perturber and
the self-gravity of the 
distorted primary. 
The differences in the mathematical
structure of these two contributions are illustrated in the Appendix.
The
transient response due to the discrete damped modes can have a strong effect
on the amplitude of the response (cf. \S3) and it is particularly important
when the modes are weakly damped with damping timescales longer than
the fly-by timescale. In this case, the 
discrete damped modes produce long-lived features which completely determine
the amplitude and the structure of the response long
after the pericenter passage of the perturber.  
Once the above algebraic equation has been solved, the calculation  of
response potential and density is straightforward
\begin{eqnarray}
\Phi_1^{resp}&=&{1 \over 2}
\sum_{lm}\left[Y_{lm}(\theta,\phi)\sum_i
a_i^{lm}u_i^{lm}(r)\right. \nonumber \\
& & \left. +Y_{lm}^*(\theta,\phi)\sum_i a_i^{lm*}u_i^{lm}(r)\right]
\end{eqnarray}
\begin{eqnarray}
\rho_1^{resp}&=&{1 \over 2}
\sum_{lm}\left[Y_{lm}(\theta,\phi)\sum_i
a_i^{lm}d_i^{lm}(r) \right . \nonumber \\
& & \left. +Y_{lm}^*(\theta,\phi)\sum_i a_i^{lm*}d_i^{lm}(r)\right]
\end{eqnarray}
where $d_i$ and $u_i$ are the bi-orthogonal density-potential basis function
and $Y_{lm}(\theta,\phi)$ are the spherical harmonics.
\section{Perturbation of dark halos}
Features excited 
in the halo can give rise to visible distortions in an embedded
disk even though dark halo perturbations are not directly observable. 
While several other processes could be responsible for the observed
distorted morphologies of stellar disks (see e.g. Sellwood \&
Merritt 1994, Levine \& Sparke 1998, Bertin \& Mark 1980, Nelson \&
Tremaine 1995), 
the role played by a distorted
galactic dark halo in exciting peculiar features in disks has been
advanced for the case of an interaction of a galaxy
with a  satellite companion by Weinberg (1995,1998b) and applied to the case
of the Milky Way-Magellanic Cloud system.
Halo deformations  leading to morphological
and kinematical lopsidedness has  been invoked by Swaters et
al. (1998) in an  analysis of two kinematically lopsided spiral
galaxies (DDO 9, NGC 4395). The ubiquity of Magellanic-class dwarf
companions of normal spirals (e.g. Zaritsky \& White 1994 and
Zaritsky et al. 1997) suggests  that such interactions may be
important sources of structure. Unbound encounters can have similar
effect as we will illustrate here.

Our standard dark halo model is an isotropic
King model with a dimensionless central potential $W_0=3.0$
(concentration \footnote{ 
the logarithm of the
ratio of the total, $R_t$, to core radius, $R_c$} $c=0.67$) a total
mass $M=6 \times 
10^{11} M_{\odot}$ and a total radius $R_t=200$ kpc. These values
are appropriate for the Galactic halo (see
e.g. Kochanek 1996) and together with a standard exponential disk
results in a fair fit to the observed rotation curve.
Given the primary model, the perturber's velocity $V$ and pericenter 
$p$  determine the structure of the response to the
perturbation. The amplitude of the response scales with the mass of
the perturber.   

 Table 1 summarizes the values of $V$ and $p$ explored here
 along with the corresponding values of the
ratio of the pericenter to the half-mass radius of the primary system,
$p/R_h$, and the ratio of the
characteristic frequency of the motion of the perturber to the
circular frequency at the edge of the primary system, $\Omega \equiv
V/p \sqrt{R_{t}^3/ GM}$. All the cases
considered here correspond to fly-bys in which part of the orbit
passes through the halo; external encounters are likely to produce
weaker distortions and will be considered later in \S 3.3.
The values of $V$ considered are chosen to cover a typical range of
 relative encounter
 velocities in the field, groups and
clusters. In order to 
clearly assess the importance of weakly damped
modes, we consider the evolution of the response both with and without
their effect for all the cases listed in Table 1.  
\begin{center}
\begin{tabular}{|c|c|c|c|}
\multicolumn{4}{c}{\bf Table 1}\\
\multicolumn{4}{c}{\bf Orbital parameters for the fly-by}\\ 
\multicolumn{4}{c}{\bf encounters
with a low-concentration}\\
\multicolumn{4}{c}{\bf primary system}\\
\hline
$V$ (km/s)& $p$ (kpc) & $p/R_h$ & $\Omega$\\
\hline
200&53.6&1.0&6.45\\
500&53.6&1.0&16.1\\
1000&53.6&1.0&32.3\\
200&107.2&2.0&3.23\\
500&107.2&2.0&8.1\\
1000&107.2&2.0&16.1\\
\hline
\end{tabular}
\end{center}

For cases that include the weakly damped modes, we must first determine
 their natural frequencies. We used the method adopted in
Weinberg (1994):
 the determinant of the dispersion matrix ${\bf D}(s)$ (see Appendix
for its definition) is evaluated on a grid of values of $s$ in the
upper ($Re (s)>0$) complex plane where the determinant is defined and
then rational functions  are used to 
perform the analytical continuation to the lower complex plane where
the zeros of the determinant corresponding to damped modes are located. 
Rational function fits of the elements of the
dispersion matrix have been adopted also for the calculation of the
residue of the inverse of the dispersion matrix necessary to include
the effects of damped modes in the response coefficients (see eq. \ref{dm}).
The characteristic damping times for the most weakly damped modes,
 $\tau_l$, are 
$\tau_1/t_{dyn}\simeq 150$ and $\tau_2/t_{dyn}\simeq 30$ for the
dipole and the quadrupole mode respectively, where $t_{dyn}$ is the
dynamical time  of the primary system defined at the half-mass
radius, $t_{dyn}\equiv \pi \sqrt{R_h^3/2GM}$.
\subsection{Importance of damped modes}
Figure 1 shows the evolution of the energy in the
perturbation, $E_p$,  as a function of $t/t_{dyn}$
(Fig.1a-b) and as a function of the
position angle of the perturber $\Theta(t)$ (Fig.1c-d) without the effects of
the damped modes. We normalize to the total background energy of the
primary system $E_{back}\equiv \nu GM^2/R_t$ ($\nu=-0.78$ for a
King $W_0=3$ model).

For a given value of the pericenter, the peak of the
response occurs after the closest approach.
The faster the perturber, the larger the value of the position angle
and the more distant the perturber location when the response of the
primary system reaches its peak.
\begin{figure*}
\plotone{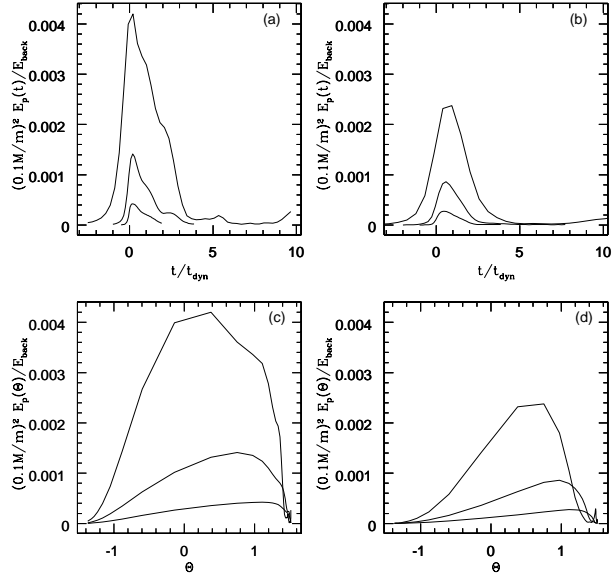}
\caption{Run of  energy in the perturbation with time  and with 
the position angle
of the perturber for fly-bys with $p/R_h=1.0$  (a) and (c) and
$p/R_h=2.0$ (b) and (d). The three curves shown in each frame correspond (from
the upper to the lower curve) to
$V=200$ km/s, $V=500$ km/s, $V=1000$ km/s. $E_{back}$ is the total
background energy of the primary system and $t_{dyn}$ is the half-mass
dynamical time. Pericenter occurs at $t=0$.}  
\end{figure*}
The strength of response increases as the velocity 
of the perturber decreases. In Figure 2 we have plotted the maximum value
of $E_p$ as a function of the velocity of the perturber for
the two values of $p$ investigated. The response amplitude is large when
the characteristic angular frequency, $V/p$, is in low-order resonance
with primary orbital frequencies.

Figure 3 and Figure 4 show the same plots shown in Figure 1 and
Figure 2 but  with the effects of the discrete weakly damped
modes taken into account. 
The maximum values of
$E_p$ are about two orders of magnitude larger than those reached when
the effects of damped modes are not included while 
the dependence of the peak of
$E_p$ on the velocity of the perturber and on the pericenter of its
orbit are similar to each other in the two cases. {\it The inclusion
of the damped modes is critical to the
persistence and the strength of the perturbation}. 
 
A simple empirical function of the form $K/V^{\alpha}$ can be used to
obtain a satisfactory fit of the curves shown in Figure 2 and Figure
4. These relations can be used to predict ensemble properties (see \S
3.4 for an example).
Table 2 summarizes the values of the parameters $K$ and $\alpha$ for
the best fits for all the cases investigated and shown in Figure 2 and
Figure 4.
\begin{center}
\begin{tabular}{|c|c|c|}
\multicolumn{3}{c}{\bf Table 2}\\
\multicolumn{3}{c}{\bf Best fit parameters for the scaling}\\ 
\multicolumn{3}{c}{\bf of $\max (0.1M/m)^2E_p/E_{back}$ with the}\\
\multicolumn{3}{c}{\bf
relative velocity of encounters}\\
\hline
Parameters&$K$ & $\alpha$\\
\hline
$p/R_h=1.0$, no-damped&0.005&1.41\\
$p/R_h=2.0$, no-damped&0.002&1.33\\
$p/R_h=1.0$, with damped &1.0&1.96\\
$p/R_h=2.0$, with damped &0.5&1.9\\
\hline
\multicolumn{3}{l}{\small The scaling of $\max
(0.1M/m)^2E_p/E_{back}$ with the relative}\\
\multicolumn{3}{l}{\small velocity of encounters, $V$, has been fitted
using the function  
}\\
\multicolumn{3}{l}{\small $f(V)={K\over (V (\hbox{km/s})/200)^{\alpha}}$.}\\ 
\hline
\end{tabular}
\end{center}
The scalings obtained will not extend to the
very low-velocity regime where they would lead to a divergence of the
energy associated to the response. In the limit of very slow
encounters we expect the energy of the response to increase more
slowly and eventually to converge to a constant value (see  Murali
\& Tremaine 1998 for an investigation of the response of galactic
halos to adiabatic perturbations).
\begin{figure*}
\plotone{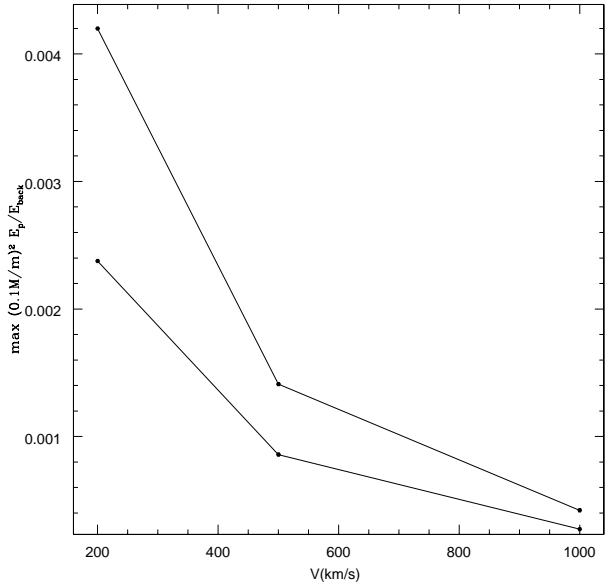}
\caption{Maximum value of the energy in the perturbation, $E_p$,
as a function of the perturber velocity $V$. The upper curve
corresponds to fly-bys with $p/R_h=1.0$ and the lower curve to fly-bys
with $p/R_h=2.0$.}
\end{figure*}

Figure 5 illustrates the differences between the response with
and without the
damped modes by separately showing the energy in the dipole and
quadrupole components for
$V=500$ km/s and $p/R_h=1.0$. As expected, the dipole 
response is significantly stronger than that of quadrupole. 
The inclusion of the damped modes leads to a much
longer and stronger dipole response, while producing no significant
difference in the quadrupole response.  
Such long-lived features may
explain the numerous cases of galaxies with distorted morphologies but
without any apparent interacting system in their vicinities
(e.g. Richter \& Sancisi 1994). For example, for a system with mass
and 
radius equal  to those of our fiducial system ($M=6\times 10^{11}
M_{\odot}$, $R=200$ kpc), the damping time of the dipole mode
($\tau_1=5\times 10^{10}$ yrs) is longer the Hubble time and any
perturbation excited 
in a dark halo during a fly-by as well as its effect on an embedded
disk will persist well after the encounter.
\begin{figure*}
\plotone{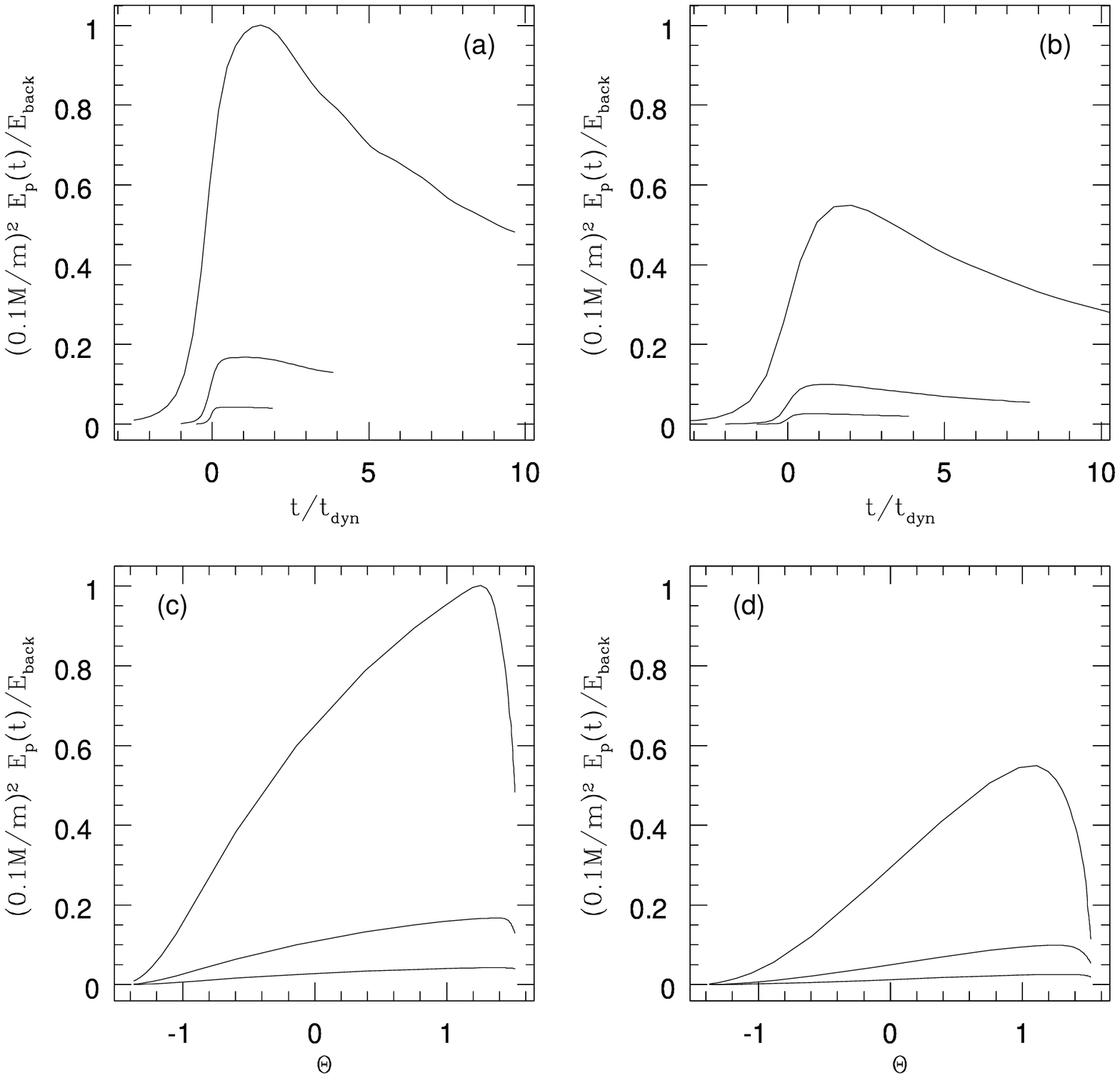}
\caption{Same as Fig. 1a-d but including the effects of
weakly damped modes.}
\end{figure*}
\begin{figure*}
\plotone{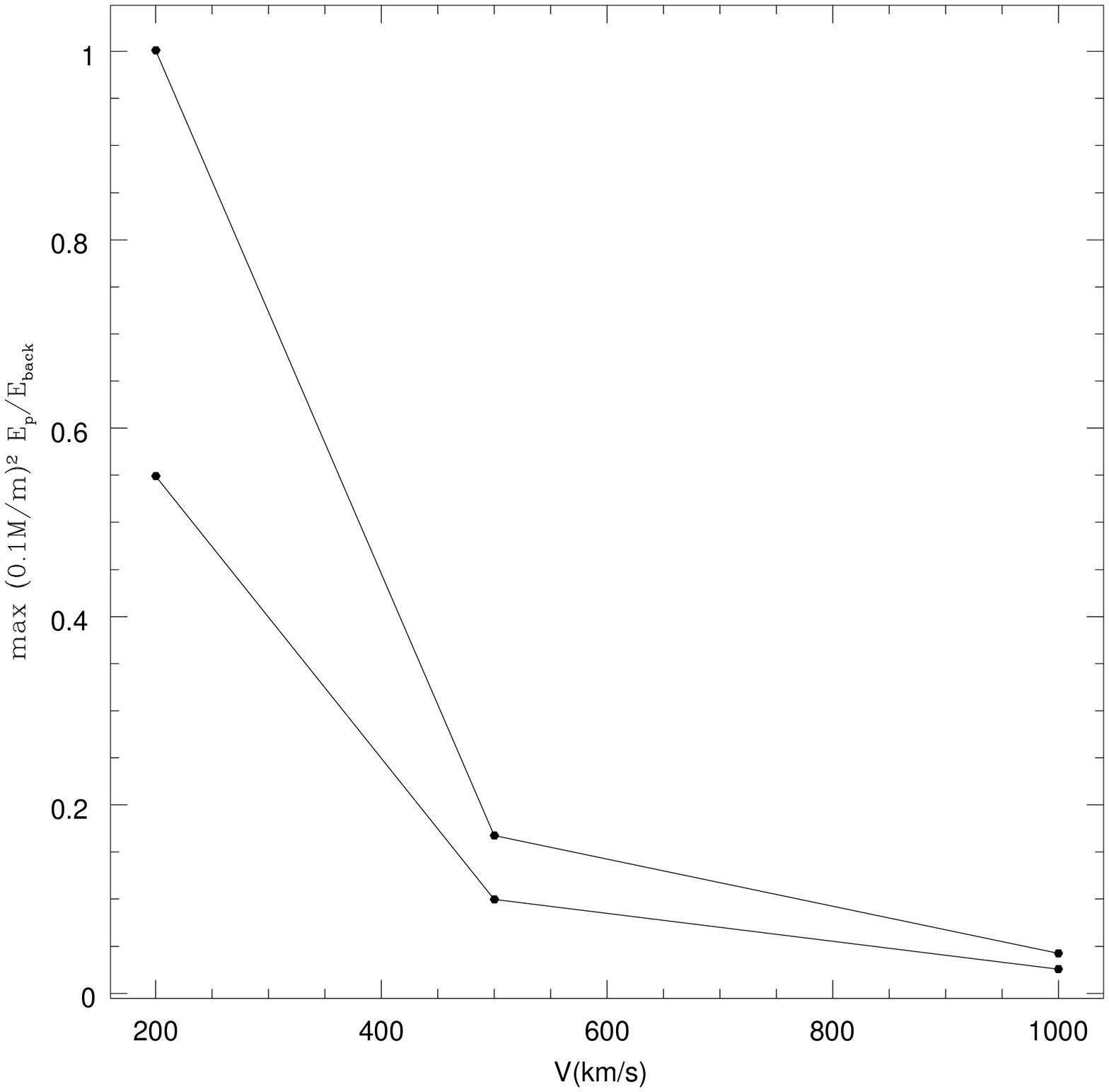}
\caption{Same as Fig. 2 but including the effects of
weakly damped modes.}
\end{figure*}
\subsection{Structure of response}
Contour plots in Figures 6 and 7 describe the density perturbation in
the orbital plane for the dipole (upper panels) and quadrupole (lower
panels) responses without and with the damped modes, respectively. 
\begin{figure*}
\plotone{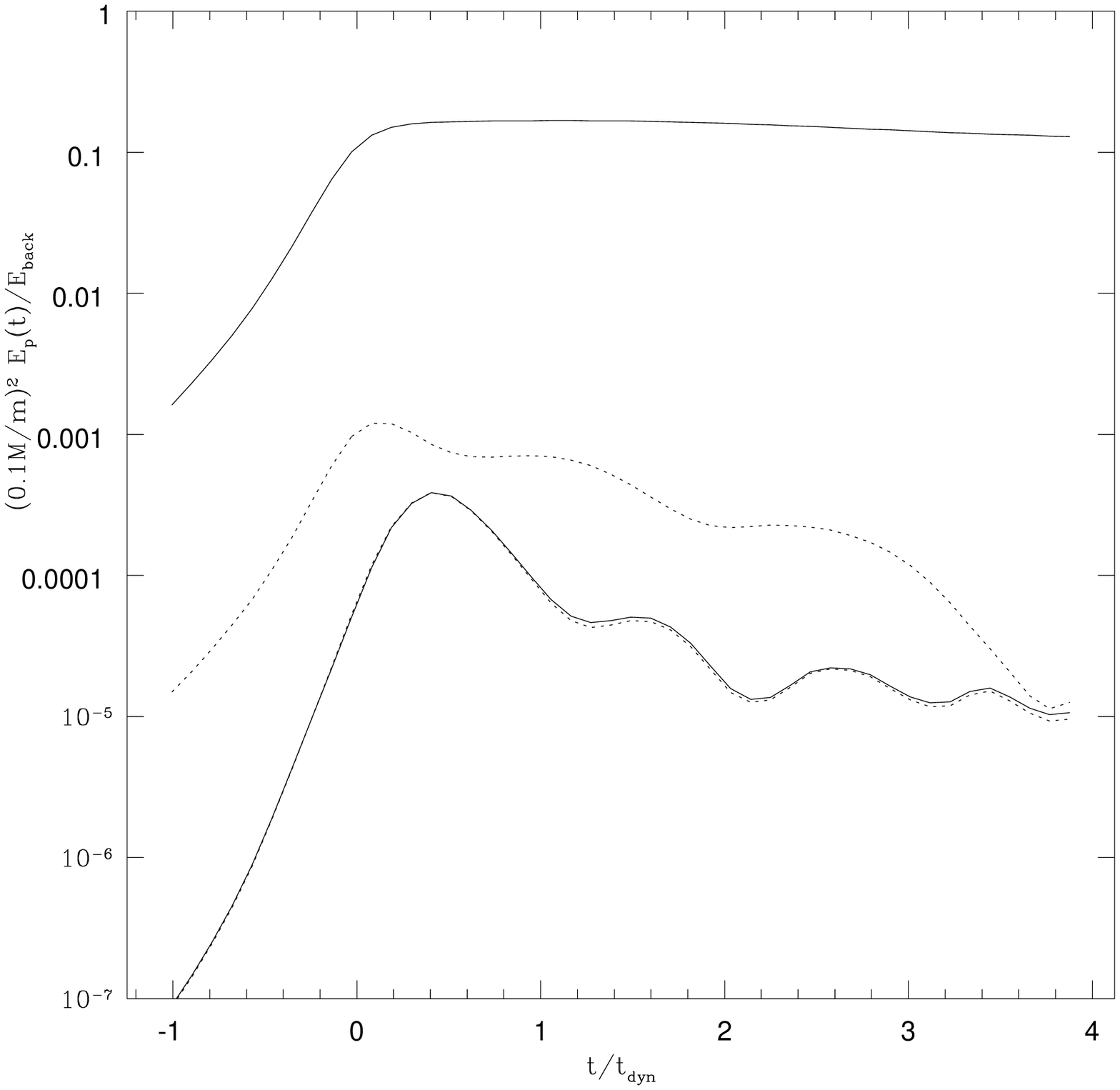}
\caption{Time evolution of the perturbation energy for a fly-by
with $V=500$ km/s, $p/R_h=1.0$. Two cases with (solid lines) and
without (dashed lines) damped modes are shown for the dipole response
(upper two curves) and quadrupole response (lower two curves).}
\end{figure*}
These cases have  $V=200$ km/s and $p/R_h=1.0$ and 
show the response of the primary galaxy at three different
points of the perturber trajectory. A lag between the orientation of the  wake
and the position angle of the perturber is evident for all velocities.
This lag is a consequence of angular momentum transfer between the
wake and the perturber  and 
tends to zero only when there is no resonance affecting the
response of the primary system \footnote{The progressive alignment between the
orientation of the wake and the position angle of the perturber for
very slow encounters is shown in Figure 12 and discussed in \S
3.3 for external encounters.}. Physically, resonant momentum exchange viewed at
the orbit level has a corresponding view as a global response:
the global response must lag the perturber in order to apply the
torque responsible for the exchange. 

Figures 8a and 8b show the ratio, $R_p/R_h$, of the radial position of
the peak of the density of the perturbation, $R_p$, to the half-mass
radius, $R_h$,  as a function of the position angle of
the perturber without the effects of weakly damped modes for $V=200$
km/s and $p/R_h=1.0,2.0$ (Fig.8a) and for 
$p/R_h=1.0$ and $V=200,500,1000$ km/s (Fig.8b). 
\begin{figure*}
\plotone{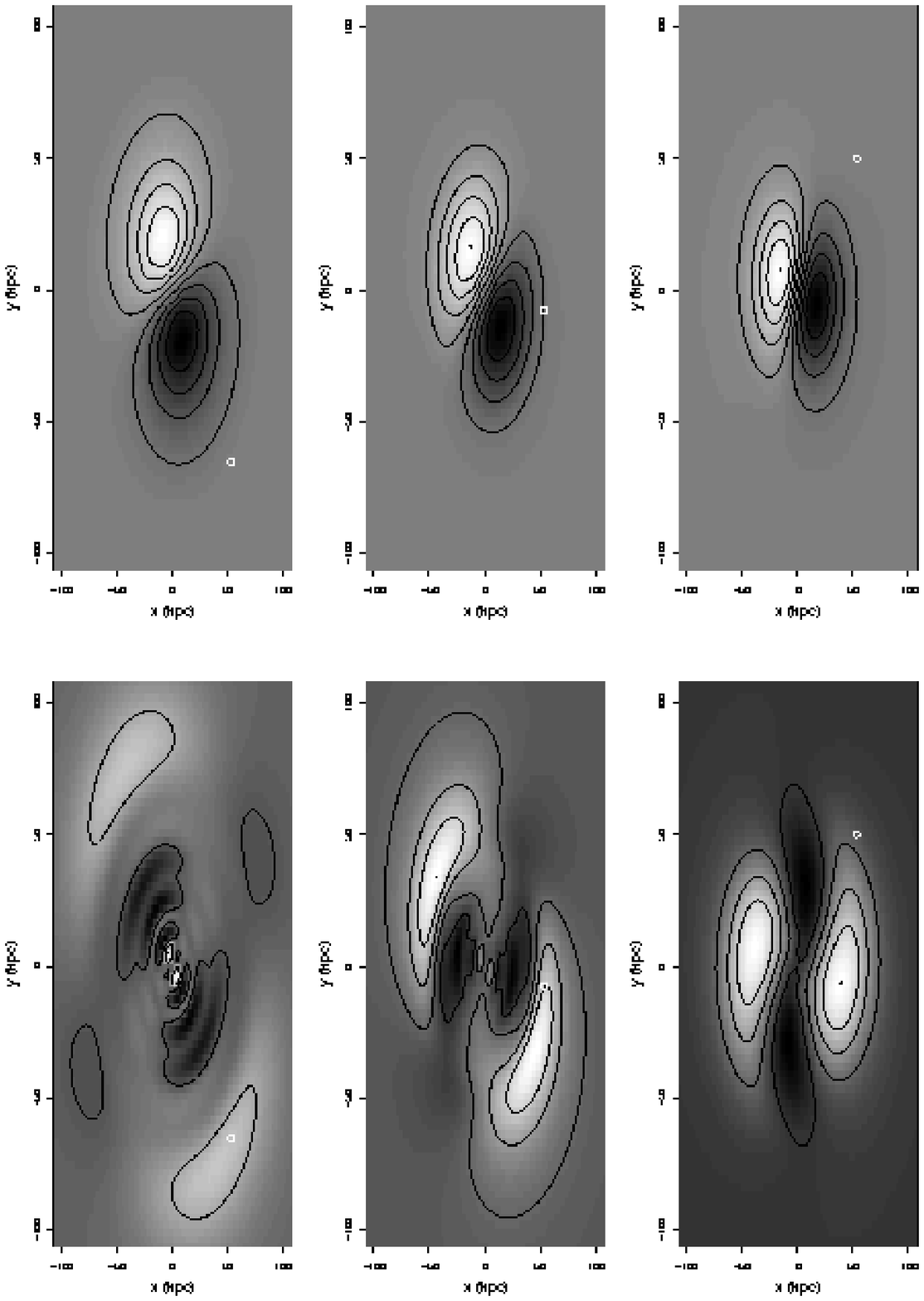}
\caption{Density distortion of the primary galaxy at three phases of
the fly-by with $V=200$ 
km/s and $p/R_h=1.0$ without the effects due to the damped
modes. Upper panels show the dipole response; lower panels show the
quadrupole response. The white dot shows the position of the
perturber. White and black regions correspond to overdensity and
underdensity respectively.}
\end{figure*}
\begin{figure*}
\plotone{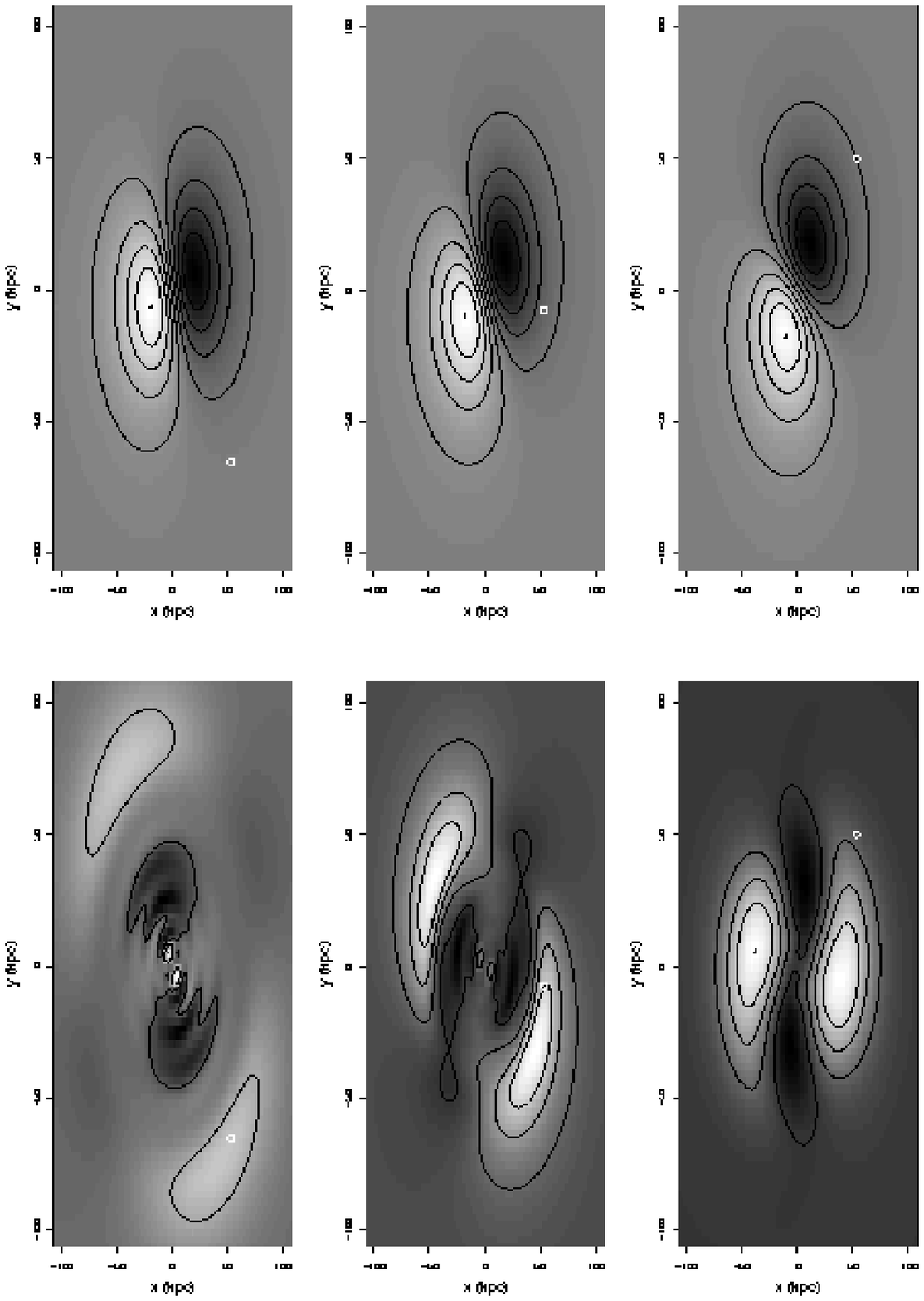}
\caption{Same as Fig. 6 but including the effects of weakly damped
modes.}
\end{figure*}
The evolution of $R_p/R_h$ is only weakly dependent on
the velocity of the perturber and on the pericenter of the orbit of
the perturber. The perturbation is efficiently transmitted to the
inner parts of the primary system and the peak of the
density of the perturbation is located 
 at $0.3\ltorder(R_p/R_h) \ltorder0.4$ or in terms of the core radius, $R_c$,
$0.4\ltorder(R_p/R_c)\ltorder 0.5$ (for almost the entire duration of
the fly-by).  The weak dependence of the
response on perturber parameters is due to the dominant effect of the
lowest-order point modes on the structure of the dispersion matrix
elements. For an analogy, consider the similarity of
bell's ring under a variety of strikes. The strong oscillations of
$R_p/R_h$ in the final phases of the fly-by are due to numerical
difficulty in determining the density peak as the response weakens.

Figures 9a and 9b show the evolution of the maximum value of the 
density of the perturbation $\delta \rho_{max}$ (normalized to the
central density of the equilibrium model) for the same cases
shown in Figures 8a and 8b. The 
maximum response obtains for slower fly-bys (cf. Fig. 1) and, for a given
value of $V$, $\delta \rho_{max}$ increases as
$p$ decreases. 
Figures 8c and 8d and 9c and 9d show the same plots as in Figures 8a-b  and
Figures 9a-b  for
fly-bys including the weakly damped modes. 
\begin{figure*}
\plotone{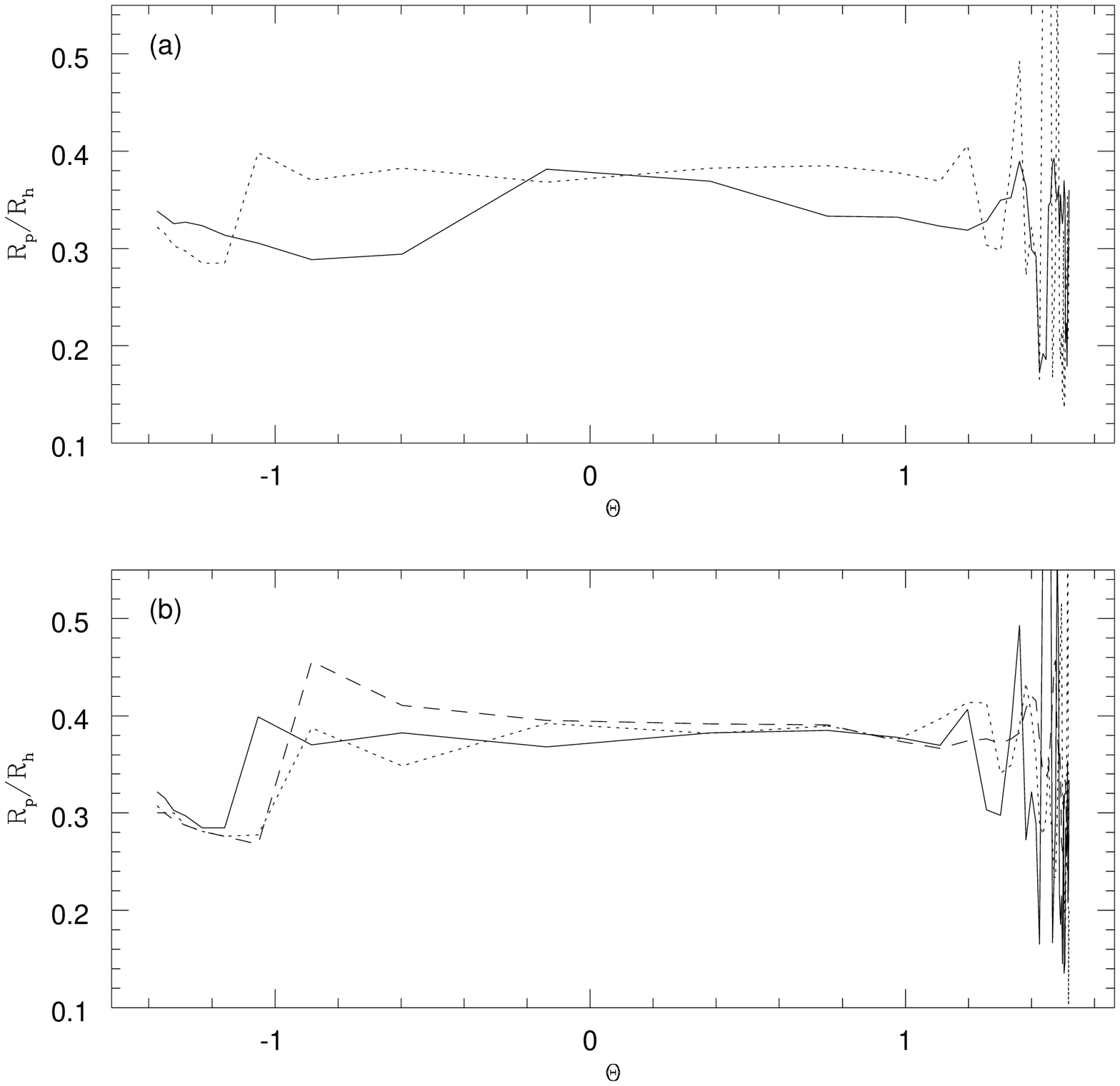}\plotone{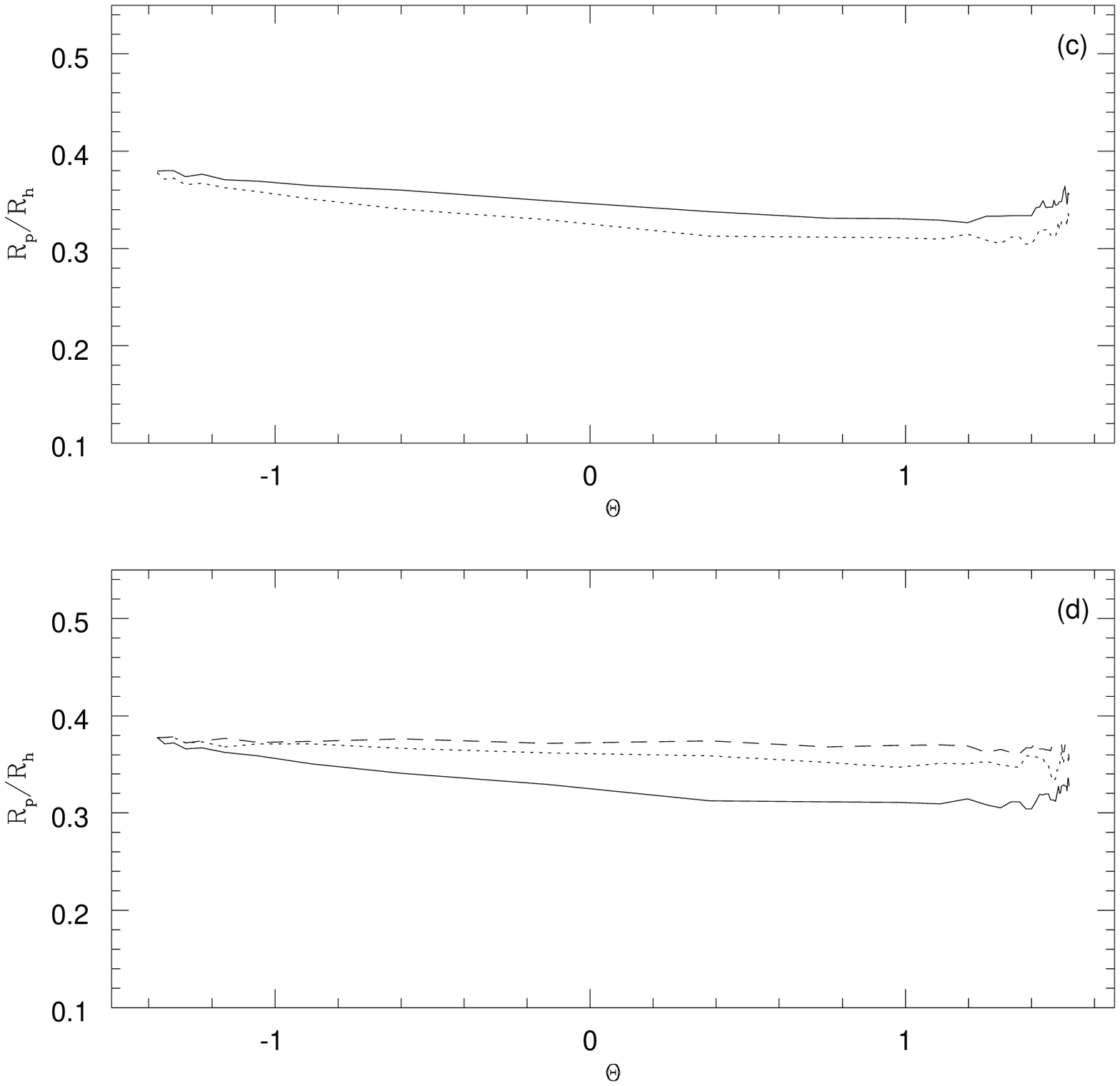}
\caption{Evolution of the ratio of the position of the peak of the
density distortion to the half-mass radius of the primary galaxy as a
function of the position angle of the perturber 
for the fly-bys with (a) $V=200$ km/s and $p/R_H=1.0$ (dashed line) and
$p/R_h=2.0$ (solid line) and (b) for the flybys with  $p/R_h=1.0$ and
$V=200$ km/s (solid line), $V=500$ km/s (short dashed line) $V=1000$
km/s (long dashed line). (c) and (d) same as (a) and (b) but including
the effects of weakly damped modes.
}
\end{figure*}
\begin{figure*}
\plotone{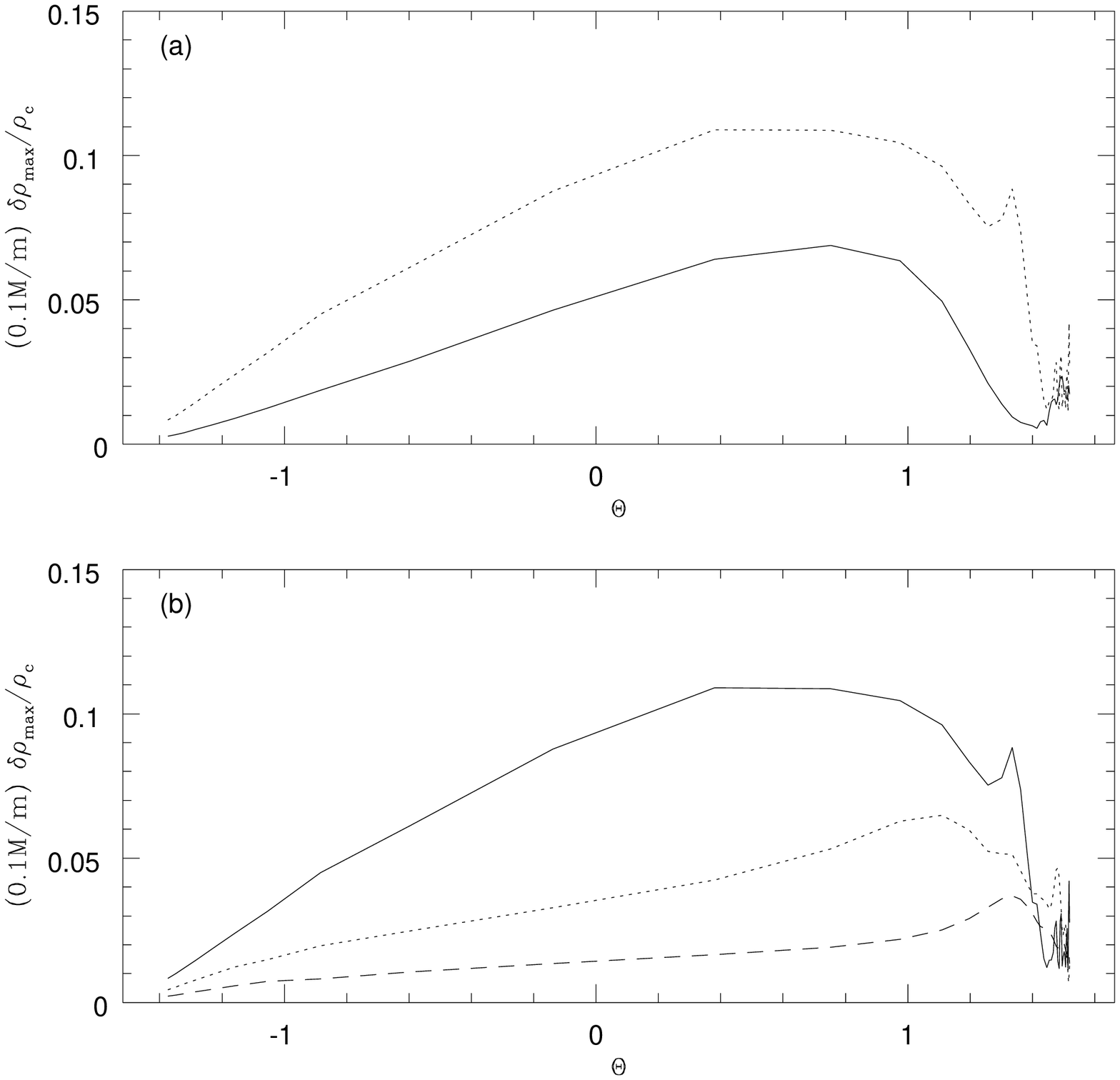}\plotone{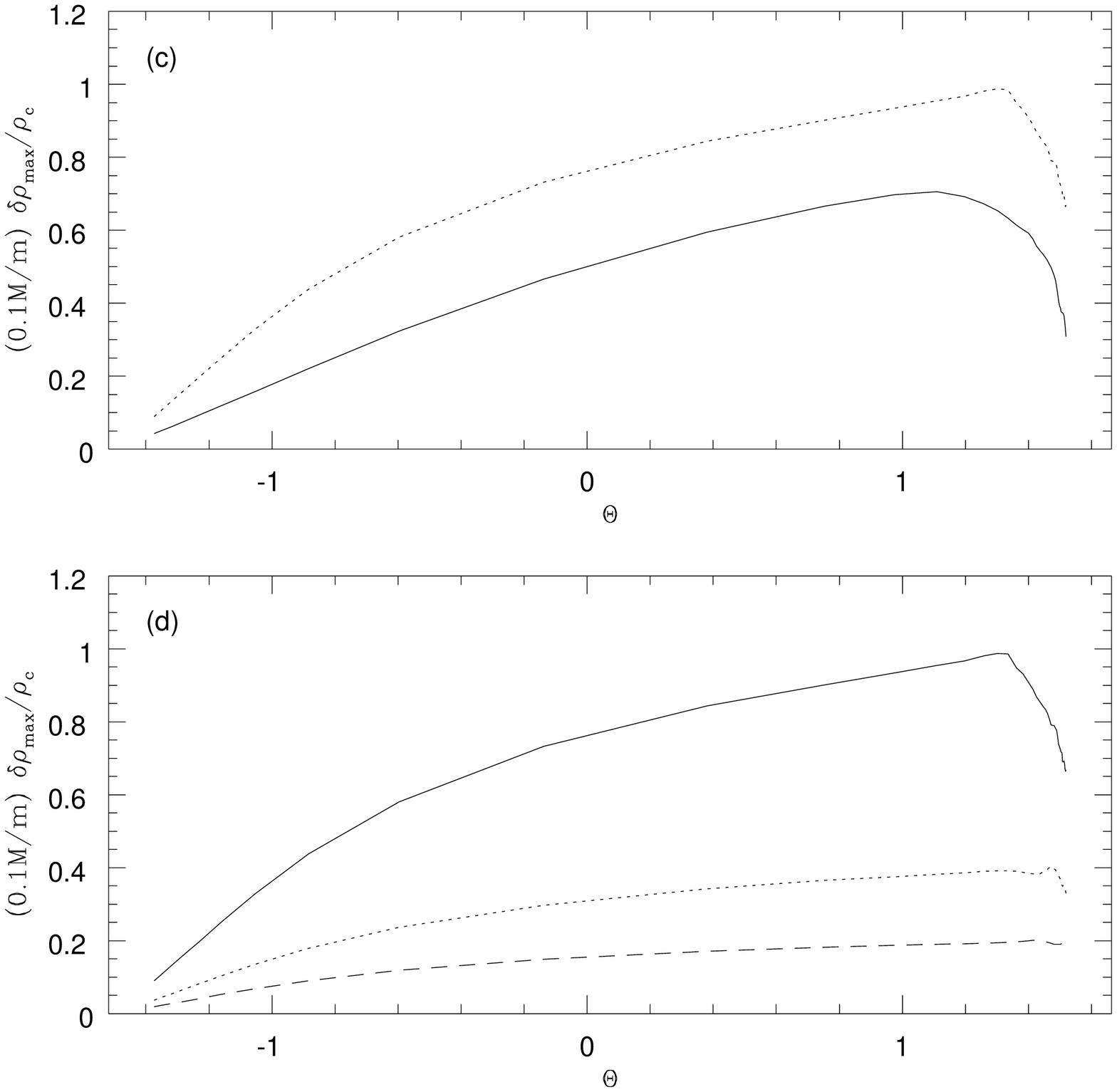}
\caption{Maximum value of the density distortion 
as function of the position angle 
for the fly-bys with (a) $V=200$ km/s and $p/R_H=1.0$ (dashed line) and
$p/R_h=2.0$ (solid line) and  for the flybys with (b) $p/R_h=1.0$ and
$V=200$ km/s (solid line), $V=500$ km/s (short dashed line) $V=1000$
km/s (long dashed line).(c) and (d) same as (a) and (b) but including
the effects of weakly damped modes.}
\end{figure*}
The position
of the peak of the density is similar to that obtained in the case
without the weakly damped modes but there are evident differences in
the time evolution of $R_p/R_h$ particularly at the beginning and at
the end of the fly-by due to the persistence of the damped modes.
The response is significantly stronger when the
effects of the damped mode are included (compare Fig.9a-b with Fig.9c-d)
\begin{figure*}
\plotone{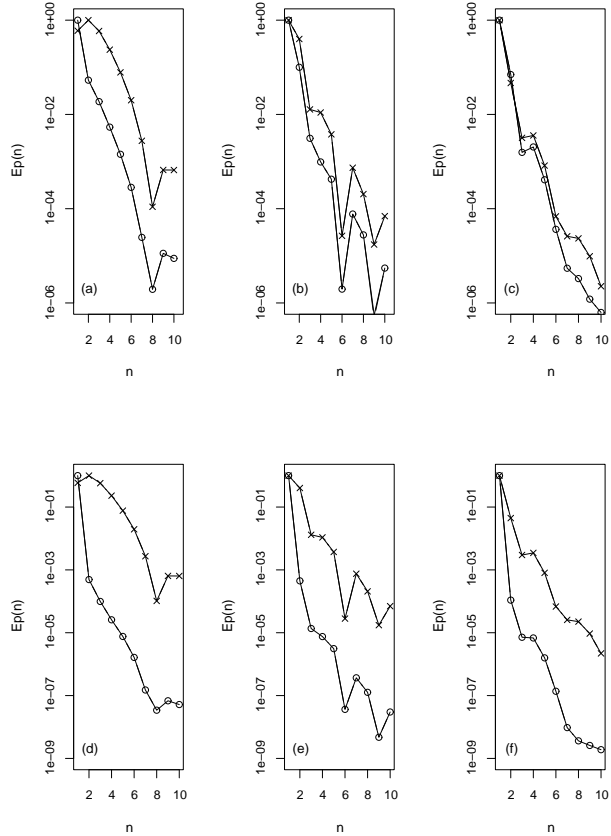}
\caption{Energy in the perturbation as a function of the orthogonal function
indices. Upper panels (a-c) correspond to the fly-by and the phases shown in
Fig.6; lower panels (d-f) correspond to the fly-by and the phases shown in
Fig.7. Open circles and crosses show the energy in
the dipole response and in the quadrupole response respectively. Each curve
is normalized to the maximum value of  $E_p(n)$ .}
\end{figure*}

Figure 10 shows the energy 
in the perturbation for the first ten basis functions and indicates the
distribution of energy with spatial scale.  
Plots shown in Figure 10 are for the
same cases and phases shown in Figure 6 and Figure 7. 
Figure 10 confirms what is already
qualitatively evident in Figure 6 and Figure 7: the energy in the dipole
response is preferentially deposited on large scales and the quadrupole
response  energy deposited on relatively smaller scales.
The typical scale represented by each index $n$ can be seen from the
plots of Figure 11 which show the density basis functions for
$l=1$.
\begin{figure*}
\plotone{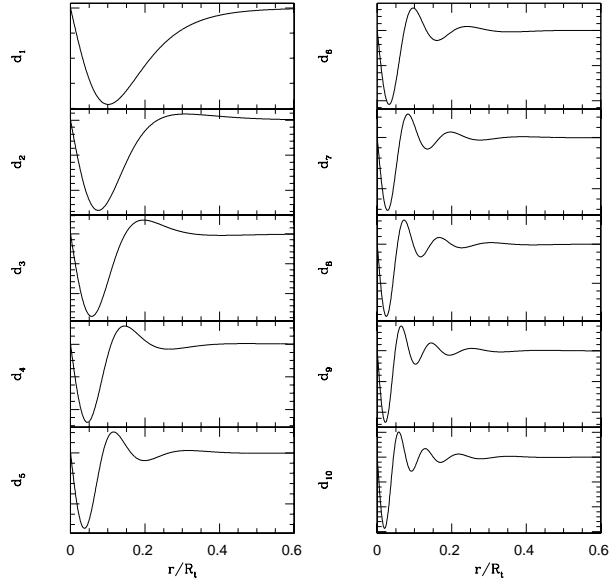}
\caption{Density basis functions used to represent 
the perturbation density (see \S 2).}
\end{figure*}
\subsection{External encounters}
External encounters, those with trajectories everywhere outside the
primary, are more frequent in groups and clusters but less dramatic.
For external encounters, the stronger dipole response corresponds to a
simple shift of the center of mass of the system leaving the weaker quadrupole
as dominant aspherical contribution. Here we briefly describe the features of
externally excited responses.

Table 3 summarizes the characteristics of the fly-bys we have
investigated. An external encounter has no intrinsic length scale and 
the frequency parameter $\Omega$ alone now determines the structure of
the response. 
The Table reports the
velocity of each fly-by assuming the pericenter to be equal
to twice the total radius of the primary system (400 kpc) and the
maximum values of the ratio of the energy in the 
perturbation to the total energy of the equilibrium model. This values
scales as shown for other values.
\begin{center}
\begin{tabular}{|c|c|c|}
\multicolumn{3}{c}{\bf Table 3}\\
\multicolumn{3}{c}{\bf External fly-bys}\\ 
\hline
$\Omega$& $V_{p=2R_t}$ (km/s)& $\max \left({1.0\over R_t/p}\right)^6\left
({1.0\over m/M}\right)^2
{E_p \over E_{back}}$ \\
\hline
0.1&23&$1.3\times 10^{-3}$\\
0.5&115&$1.3\times 10^{-3}$\\
1.0&231&$7.0\times 10^{-4}$\\
2.5&578&$4.7\times 10^{-4}$\\
5.0&1150&$2.8\times 10^{-4}$\\
\hline
\multicolumn{3}{l}{\small $V_{p=2R_t}$ is the relative velocity of the
encounter for }\\
\multicolumn{3}{l}{\small the given value of $\Omega$  reported
in the first column for}\\ 
\multicolumn{3}{l}{\small for $p=2R_t=400$ kpc.}\\  
\hline
\end{tabular}
\end{center}

One sees that the effects of
external encounters with less massive perturbers are weak and
they can not produce a significant deformation in the primary system.
On the other hand, if the perturber is more massive than the perturbed system
even external fly-bys can be very important in affecting the structure
of a galaxy. For example, in order to have $\max(E_p/E_{back})\simeq
0.01$ for an  encounter with $p=2R_t$ the ratio of the mass of the
perturber to that of the perturbed must be 
$m/M\simeq 30$. These encounters are  relevant for spiral
galaxies in clusters where they can undergo frequent 
external encounters with more massive giant
ellipticals. A recent investigation by Rubin et al. (1999) has shown
that about half of a sample of 89 spirals in the Virgo cluster exhibit signs of
kinematic disturbances in their rotation curves; according to that
investigation disturbed galaxies are preferentially on radial orbits
likely to cross the central regions of the clusters 
where such encounters  are
likely to occur.  

The five panels of Figure 12 show
the contour plots of the density of the response when $E_p/E_{back}$
reaches its maximum value. As described in \S 3.2, the response reaches its maximum strength
at different phases of the 
orbit of the perturber depending on the value of $\Omega$: the larger
the value of $\Omega$, the more distant the perturber is from the
pericenter of its 
orbit when the response is maximum. As the characteristic frequency of
the fly-by decreases and the 
perturbation tends to the adiabatic regime,
individual resonances cease to be relevant in determining the
structure of the response
and the orientation of the wake aligns with the position of the perturber.
Finally, we  note that 
some features are present at small radii and these may cause
visible distortions in an embedded disk.

\begin{figure*}
\plotone{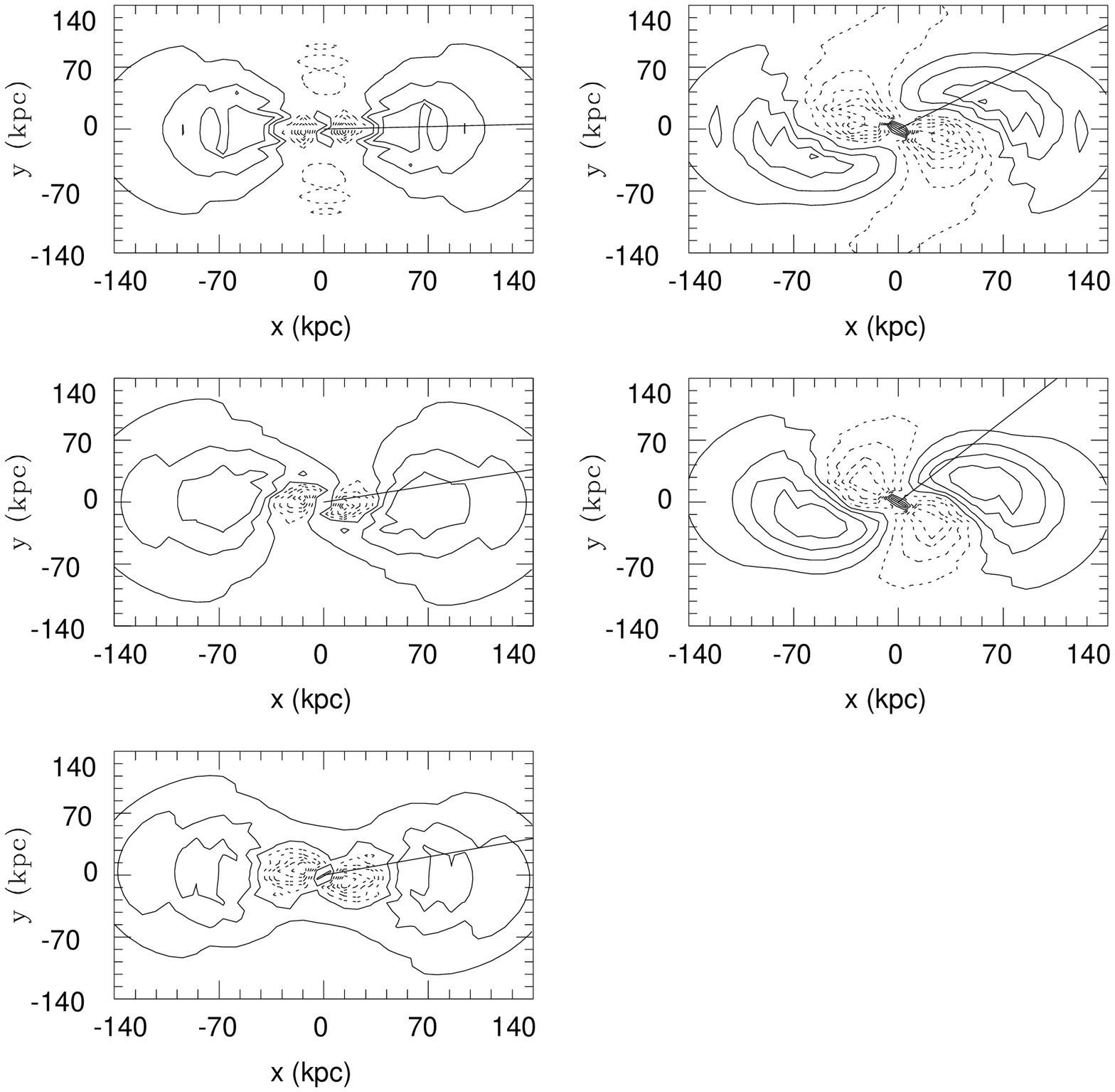}
\caption{Density distortion of the primary galaxy when the
perturbation energy is maximum. The three
frames on the left  correspond to fly-bys with (from the upper to the lower
panel) frequencies  $\Omega=0.1,~0.5,~1.0$; the two frames on the
right are for the 
fly-bys with  $\Omega=2.5,~5.0$. The straight solid line in each panel
indicates the  position angle of the perturber.}
\end{figure*}
\subsection{Summary and discussion: dark halos}
We have investigated the response of a galactic dark
halo  to the perturbation induced by external and internal fly-bys. 
External fly-bys are effective when
the mass of the perturber is larger than that of the
primary system considered.  For example,  
perturbations excited in spirals during external fly-bys with more
massive giant 
ellipticals can play an important role
in altering their structure and kinematics. Similarly, external
encounters will produce significant structure in dwarfs in the
neighborhood of normal spirals.   

We have shown that the perturbation excited
during an internal fly-by  encounter can be
efficiently transmitted to the inner regions of the halo where it
can affect the structure of an embedded stellar disk. The halo
distortions can be significant and persistent for typical dwarf interlopers.
This may lead to distorted morphologies like those observed in warped and
lopsided disks. The survey over different velocities
and pericenter distances provides a quantitative
measure of  the expected importance of the process investigated in
different environments. 

For an overall indicator of the fraction of objects distorted  as a result of 
fly-by encounters, we
have calculated the mean value of the peak relative distortion for
different environments,
\be
\langle \max
~(0.1M/m)^2{E_p/E_{back}}\rangle={\int {K\over V^{\alpha}}f(V) V^2
dV\over \int f(V) V^2 dV}
\ee
where  $f(V)$ is the distribution of the relative velocities of
encounters  which we take equal to a Gaussian 
with dispersion $\sigma$. Figure 13 shows $\langle \max
(0.1M/m)^2E_p/E_{back}\rangle$ as a function of $\sigma$ for the four
scaling laws listed in Table 2. 
\begin{figure*}
\plotone{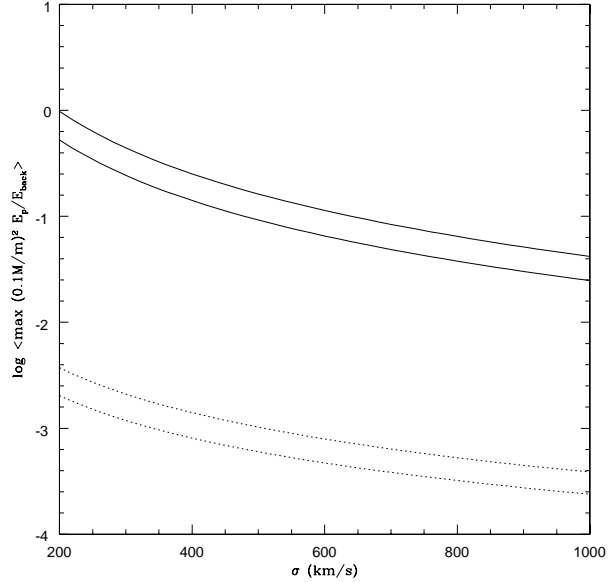}
\caption{Mean value of $\max (0.1M/m)^2E_p/E_{back}$ assuming  a Gaussian
distribution of the relative 
velocity of encounters versus the dispersion, $\sigma$, of the
distribution. Solid lines (upper line for $p/R_h=1$, lower line for
$p/R_h=2.0$)  are for runs including the effects damped
modes; dashed lines are for runs without the effects of damped modes.}
\end{figure*}
Generally, low velocity dispersion compact
groups with number densities $n \sim 10^2 \hbox{ Mpc}^{-3}$ and
velocity dispersions $\sigma \sim 10^2$ km/s are dominated by
low-velocity close encounters which lead to the strongest response in
the perturbed galaxies; cluster of galaxies which are characterized
by higher velocity dispersions ($\sigma \sim 10^3$ km/s) and lower
number densities ($n \sim 10^0-10^1 \hbox{ Mpc}^{-3}$) encounters are
expected to produce less pronounced effects.

The effects of damped modes on the strength
and the persistence of the 
features excited by the perturber are significant.
In particular the persistence of the deformations due to the
long-lived dipole mode may  explain the origin of the numerous
lopsided  disks in field galaxies which have no  apparent companion
(see e.g. Richter \& Sancisi 1994, Zaritsky \& Rix 1997). We have
separated the response due to damped modes in Figure 13 to highlight their
importance, however, true physical response includes both.
Damped modes may be artificially suppressed in n-body simulations with
small particle number owing to fluctuation-induced orbital diffusion.

In the next section, we will focus our attention on
elliptical galaxies modeled as high-concentration King models for
which the effects of discrete modes have not been considered because they
have much shorter damping times (Weinberg 1993) and thus
are not as relevant as they are for dark halos \footnote{In addition, their
inclusion presents challenging problems connected with the stability of
the numerical evaluation of the poles and the residues of the
dispersion matrix.}.  
\section{Perturbation of cluster elliptical galaxies}
\begin{center}
\begin{tabular}{|c|c|c|c|}
\multicolumn{4}{c}{\bf Table 4}\\
\multicolumn{4}{c}{\bf Orbital parameters for the fly-by}\\ 
\multicolumn{4}{c}{\bf encounters
with a high-concentration}\\
\multicolumn{4}{c}{\bf primary system}\\
\hline
$V$ (km/s)& $p$ (kpc) & $p/R_h$ & $\Omega$\\
\hline
200&7.0&0.5&17.9\\
500&7.0&0.5&44.7\\
1000&7.0&0.5&89.5\\
\hline
200&14.0&1.0&8.9\\
500&14.0&1.0&22.4\\
1000&14.0&1.0&44.7\\
\hline
200&42.0&3.0&3.0\\
500&42.0&3.0&7.5\\
1000&42.0&3.0&14.9\\
\hline
200&97.8&7.0&1.28\\
500&97.8&7.0&3.2\\
1000&97.8&7.0&6.4\\
\hline
\end{tabular}
\end{center}
We will model the primary
system by an isotropic King model with $W_0=7.0$ which has a concentration
$c=1.53$ \footnote{A core-free primary increases the cost of the
computation without significant changes in the response beyond the core
region}. We
adopt typical mass, total radius and half-mass radius values for cluster
elliptical galaxies:  
$M=10^{12} M_{\odot}$ and $R_t=120$ kpc, $R_h=14.0$ kpc.
Table 4 summarizes the properties of our model grid (see Table 1 for
parameter definitions).  
The overall approach and procedure otherwise follows that described in
\S 3 with similar physical interpretation. \S 4.1 briefly presents the
response diagnostics 
developed in \S 3, highlighting the differences between responses in
the low- and high-concentration primaries. \S 4.2 explores
observational diagnostics of the fly-by induced 
asymmetries and proposes a generalization of one of the standard
observational 
asymmetry parameters to test the dynamical mechanism explored here.
 
\subsection{Structure of response}
Figure 14 shows the time evolution of the perturbation energy. As
expected (cf. \S 3), the   
innermost fly-bys induce the strongest response and,
for a given value of the pericenter, the 
response becomes weaker as the velocity of the perturber increases. The time
evolution of $E_p$ is characterized by several peaks and
a non-negligible amplitude even as $\Theta$ tends to $\pi/2$
and the perturber moves away from the primary galaxy. The
persistence of these features well past the
closest passage of the perturber may explain the origin 
of distorted morphology in galaxies without any close
visible companion.
\begin{figure*}
\plotone{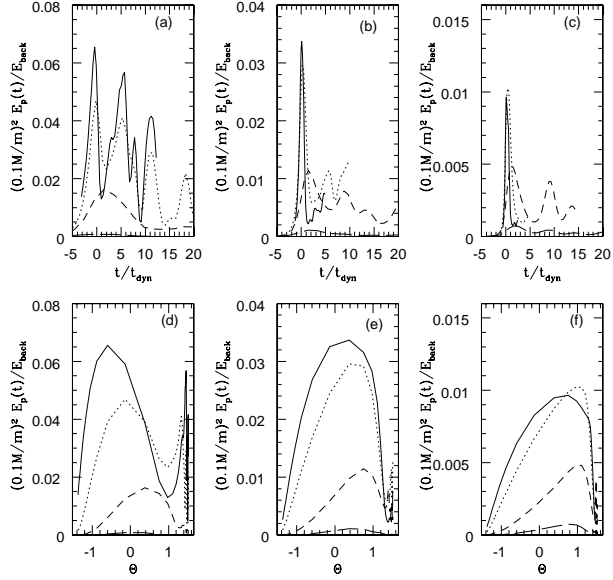}
\caption{Evolution of the perturbation energy as a function of time
($t=0$
is the time of passage of the perturber at the pericenter of its
orbit) (a-c) and the position angle of the perturber
(d-f). Panels (a) and (d) are for fly-bys with $V=200$ km/s, panels
(b) and (e) for $V=500$ km/s, panels (c) and (f) for $V=1000$ km/s. In
each panel the curves shown correspond to different values of the
pericenter of the orbit of the perturber: $p/R_h=0.5$ (solid line),
$p/R_h=1.0$ (dotted line), $p/R_h=3.0$ (dashed line), $p/R_h=7.0$
(long dashed line).}
\end{figure*}
Figure 15 shows the evolution of $R_p/R_h$. The response is
manifested well inside the primary galaxy and it affects the very
inner regions of the system $0.05\ltorder R_p/R_h \ltorder 0.15$
($0.2\ltorder R_p/R_c \ltorder 0.6$) even when the 
pericenter of the perturber is as large as $3R_h$ or $7R_h$.
The value of $R_p$ is determined by the dominant resonances determining
the structure  of the response of the primary system.
The larger dynamic range in primary profile enables excitation of
high-order resonances 
between the characteristic frequency of the motion of the perturber
and the frequencies of the orbital motion at smaller radii in the
primary system. The rectilinear trajectory considered here provides 
continuum spectrum of frequencies that peaks at approximately $V/p$.
\begin{figure*}
\plotone{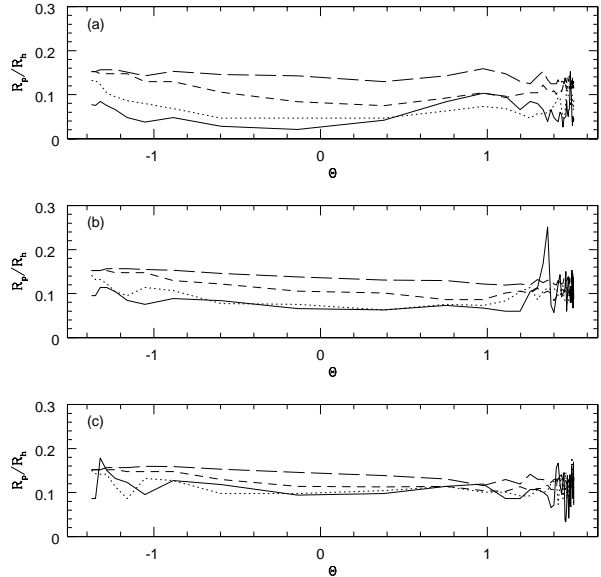}
\caption{Evolution of the position of the peak 
density distortion as a
function of the position angle  for
the fly-bys with $V=200$ km/s (a), $V=500$ km/s (b) and $V=1000$ km/s
(c). In each panel the four curves correspond to different values of the
pericenter of perturber: $p/R_h=0.5$ (solid line), $p/R_h=1.0$ (dotted
line), $p/R_h=3.0$ (dashed line), $p/R_h=7.0$ (long dashed line).} 
\end{figure*}

We found that $R_p/R_c$ is roughly constant for all the
fly-bys investigated. This agrees with
the results obtained by Weinberg (1994) in his analysis of the
structure of  weakly damped modes in isotropic King models.
This implies that more concentrated models are more
efficient in transmitting  to the inner regions
the perturbation excited during a fly-by (compare Fig.8 with Fig.15).

Figure 16 shows the 
density distortions for a fly-by with $V=200$ and $p/R_h=1.0$ at three
different phases of the orbit  of the perturber.  
As in previous cases, the maximum
response occurs well inside the primary system and the perturbation is
efficiently transmitted to the inner regions of the system far inside
the orbital pericenter.
\begin{figure*}
\plotone{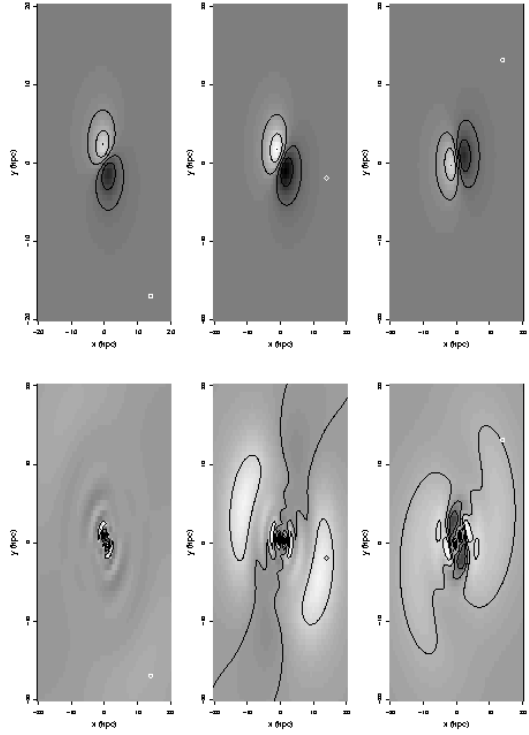}
\caption{Density distortion of the primary galaxy at three phases of
the fly-by with $V=200$ 
km/s and $p/R_h=1.0$. Upper panels show the dipole response; lower
panels show the quadrupole response. The white dot shows the position of the
perturber.}
\end{figure*}
\begin{figure*}
\plotone{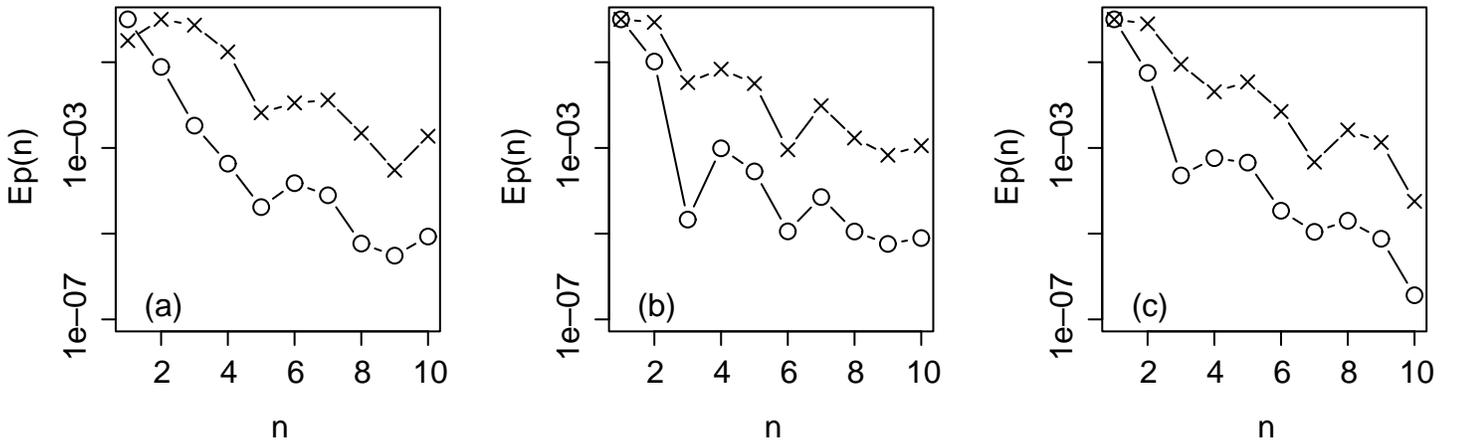}
\caption{Perturbation energy  as a function of the orthogonal function
indices for three phases shown in Figure 16. 
In each panel, open circles (crosses) indicate dipole (quadrupole) 
 response. Each curve
is normalized to the maximum value of $E_p(n)$.}
\end{figure*}
Figure 17 shows  plots of $E_p(n)$ for the same  case and phases
shown in Figure 16 describing the distribution of energy at different scales.
Similar to the results in \S 3, the dipole response is dominated by
large scale features while in the quadrupole response there is a non negligible
contribution to the total energy of the perturbation from features at
smaller scales.

\subsection{Diagnostics of asymmetry in the primary galaxy}
We employ an established asymmetry diagnostic to compare our
predictions to observations and suggest a straightforward
generalization which will help discriminate the response mechanism
considered here from others.
We will focus our attention on 
an asymmetry parameter $A$ for image data, defined as
\be
A\equiv {1\over 2} {\sum \vert  I(x,y)- I_{rot}(x,y) \vert \over
\sum I(x,y)} \label{asym}
\ee
where $I(x,y)$ and $I_{rot}(x,y)$ are the brightness at a given point
$(x,y)$ of the original image and in the image
after a 180 degree rotation  about the position of the maximum
density of the primary system, respectively.

For our calculation we use the density of the system on the orbital
plane and thus our theoretical values are not affected by any projection
effect and will provide a measure of the real asymmetry of the plane.
Observational values, on the other hand, are determined using the
projected density and part of the underlying 
asymmetry of the orbital plane is likely to be washed out. 
In order to evaluate the effect of using the
projected density on the orbital plane instead of the real plane
density, we have repeated some of the calculations\footnote{These
calculations require the evaluation of the density distortion on a
three-dimensional grid which leads to an increase in the
computational time by approximately a factor 100}: we have
found that, for a given value of $m/M$, the values of $A$ obtained using
the projected density are smaller than those obtained using the
real plane density by a constant factor 
independent of the perturber orbital parameters and approximately equal to
3.8; this implies that, using the
projected density, all the values of  $A$ (as well as those of the
parameter $S$ introduced later in this section)  and the numerical
parameters in the scaling laws of the amplitude of $A$  with the
perturber's orbital 
parameters discussed in this section are not altered but they are produced
by a perturber with a mass equal to $0.38M$ (instead of $0.1M$ adopted in
the rest of this section). 

The $A$ parameter was first introduced by Abraham et al. (1996a) to estimate
the asymmetry observed in galaxies in the Medium Deep Survey and in
the Hubble Deep Field and it has been recently adopted in several
investigations on the morphology of  galaxies in different
environments (e.g. van den Bergh et al. 1996, Naim, Ratnatunga \&
Griffiths 1997, Marleau \& Simard 1998, Conselice \& Bershady 1998,
Conselice \& Gallagher 1999). 
The values of $A$ reported in these investigations range 
from $0$ for the 
most symmetric systems to $0.5$ for those showing the
strongest asymmetries.  A comparison of the distribution
of values of $A$ for a sample of local galaxies, galaxies in the
Medium Deep Survey and in the Hubble Deep Field was reported by
Abraham et al. (1996a, 1996b). Their results show that the distribution of $A$
for galaxies in the HDF is significantly skewed toward larger values of
$A$ compared to the distribution for galaxies in the MDS which in turn
is skewed to larger values of $A$ relative to the distribution of $A$
for a sample of local galaxies. This trend is likely to be due to
the higher interaction and merging rates for distant galaxies in the HDF
and the MDS. In principle, a high value of $A$ could also
result from an asymmetric distribution of star forming regions, however a
recent investigation by Conselice \& Bershady (1998) shows that the
colors of many asymmetric objects in the HDF are not blue enough to
support this hypothesis.

In addition, we have calculated the displacement, $S$,
between the position of the peak of the density of  
the primary system and the position of its center of mass. The estimate of
$S$ is of particular interest since many
observations have provided evidence of off-center
galactic nuclei and of significant displacement of isophotal centers
(see e.g. Miller \& Smith 1992 for a short overview of 
several observational analyses).  
In particular, Naim et al. (1997)  have adopted 
distances between the centers of different isophotal curves 
as one of the diagnostic
parameters characterizing the degree of peculiarity of galaxies in a
sample of moderate-redshift systems and Mendes de Oliveira \&
Hickson (1994) have investigated the morphology of early-type
galaxies in compact groups and used the evidence of non-concentric
isophotes as a diagnostic to identify perturbed galaxies.
\subsection{Evolution of the asymmetry diagnostic parameters}
Figure 18 shows the time evolution of the asymmetry parameter $A$ defined  
above (eq. \ref{asym}). 
\begin{figure*}
\plotone{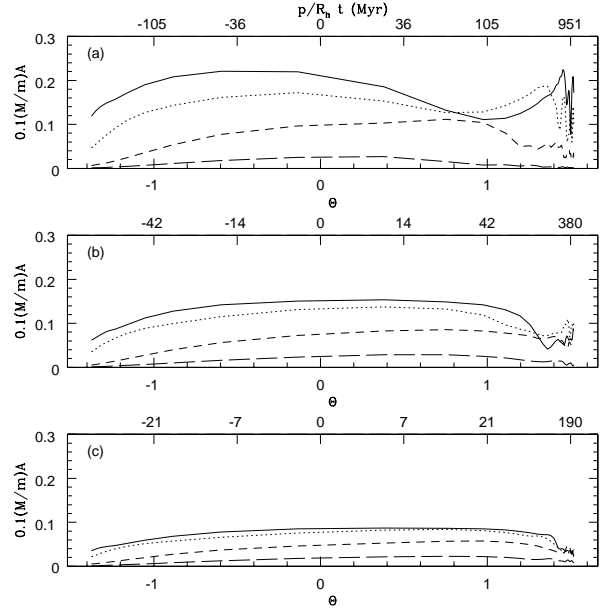}
\caption{Evolution of the asymmetry parameter $A$ (see eq.\ref{asym}) 
as a function of the position angle for
the fly-bys with $V=200$ km/s (a), $V=500$ km/s (b) and $V=1000$ km/s
(c). In each panel the four curves correspond to different values of the
pericenter of perturber: $p/R_h=0.5$ (solid line), $p/R_h=1.0$ (dotted
line), $p/R_h=3.0$ (dashed line), $p/R_h=7.0$ (long dashed line).
} 
\end{figure*}
Although several well-documented processes contribute to the observed
distorted morphologies, the majority of 
the values in the distribution of $A$ obtained for  galaxies in
the MDS and in the 
HDF by Abraham et al. (1996) (see their Fig. 2) 
fall well in the range of  values of $A$  we have obtained in our
analysis. Mergers and strong interactions with more massive
systems are likely to be responsible for the production of the extreme
values of $A$ ($A\gtorder 0.3$) observed in some galaxies.
We point out that the observational data  refer to all the
morphological types including spiral galaxies to which our
results do not directly apply.

Figure 19 shows the time evolution of $S$.
\begin{center}
\begin{tabular}{|c|c|c|c|c|}
\multicolumn{5}{c}{\bf Table 5}\\
\multicolumn{5}{c}{\bf Best fit parameters for the scaling of}\\ 
\multicolumn{5}{c}{$\max (0.1M/m)A$ \hbox{\bf and} $\max
(0.1M/m)S~(pc)$}\\
\multicolumn{5}{c}{\bf  with the
relative velocity of encounters}\\
\hline
id.&$K_A$ & $\alpha_A$&$K_S$ & $\alpha_S$\\
\hline
$p/R_h=0.5$&0.24&0.6&$2.5\times 10^3$&0.6\\
$p/R_h=1.0$&0.19&0.5&$1.9\times 10^3$&0.5\\
$p/R_h=3.0$&0.11&0.4&$0.8\times 10^3$&0.4\\
$p/R_h=7.0$&0.03&0.1&$0.1\times 10^3$&0.0\\
\hline
\multicolumn{5}{l}{\small The scaling of $\max
(0.1M/m)^2A$ and  $\max
(0.1M/m)^2S$ with}\\
\multicolumn{5}{l}{\small the relative velocity of encounters, $V$,
has been fitted using 
}\\
\multicolumn{5}{l}{\small the function $f(V)={K\over (V (\hbox{km/s})/200)^{\alpha}}$.}\\ 
\hline
\end{tabular}
\end{center}
 As expected, the trend with $V$ and $p$ is similar to that
obtained for the asymmetry parameter $A$, with slow and close
encounters being those capable of producing the largest values of $S$.
Even fly-bys with large values of $p$ are
capable of affecting the 
inner regions of the primary galaxies and lead to a significant
displacement of the position of the density peak from the center of mass
of the system.
\begin{figure*}
\plotone{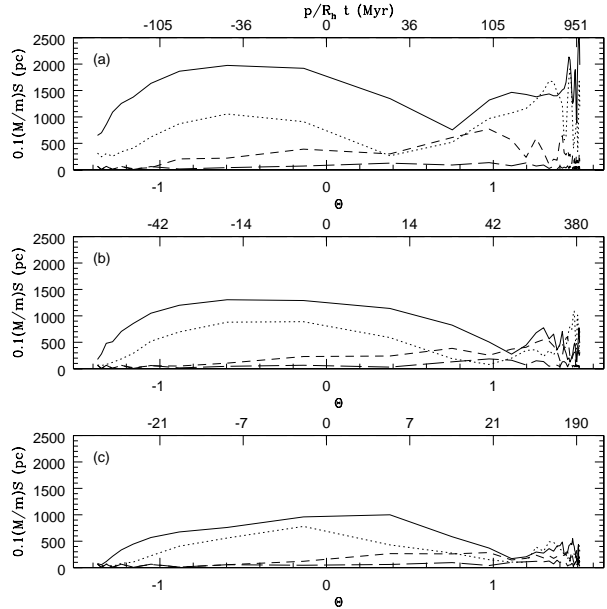}
\caption{Evolution of the distance, $S$, between the position of the
density peak and its center of mass as a function
of the position angle of the perturber for 
the fly-bys with $V=200$ km/s (a), $V=500$ km/s (b) and $V=1000$ km/s
(c). In each panel the four curves correspond to different values of the
pericenter of perturber: $p/R_h=0.5$ (solid line), $p/R_h=1.0$ (dotted
line), $p/R_h=3.0$ (dashed line), $p/R_h=7.0$ (long dashed line).}
\end{figure*}

The dependence of both asymmetry parameters on $V$ and
$p$ is summarized in Figure 20 which shows the maximum
values taken by  $A$ and $S$  as a
function of $p/R_h$ for the three values of $V$ considered.
\begin{figure*}
\plotone{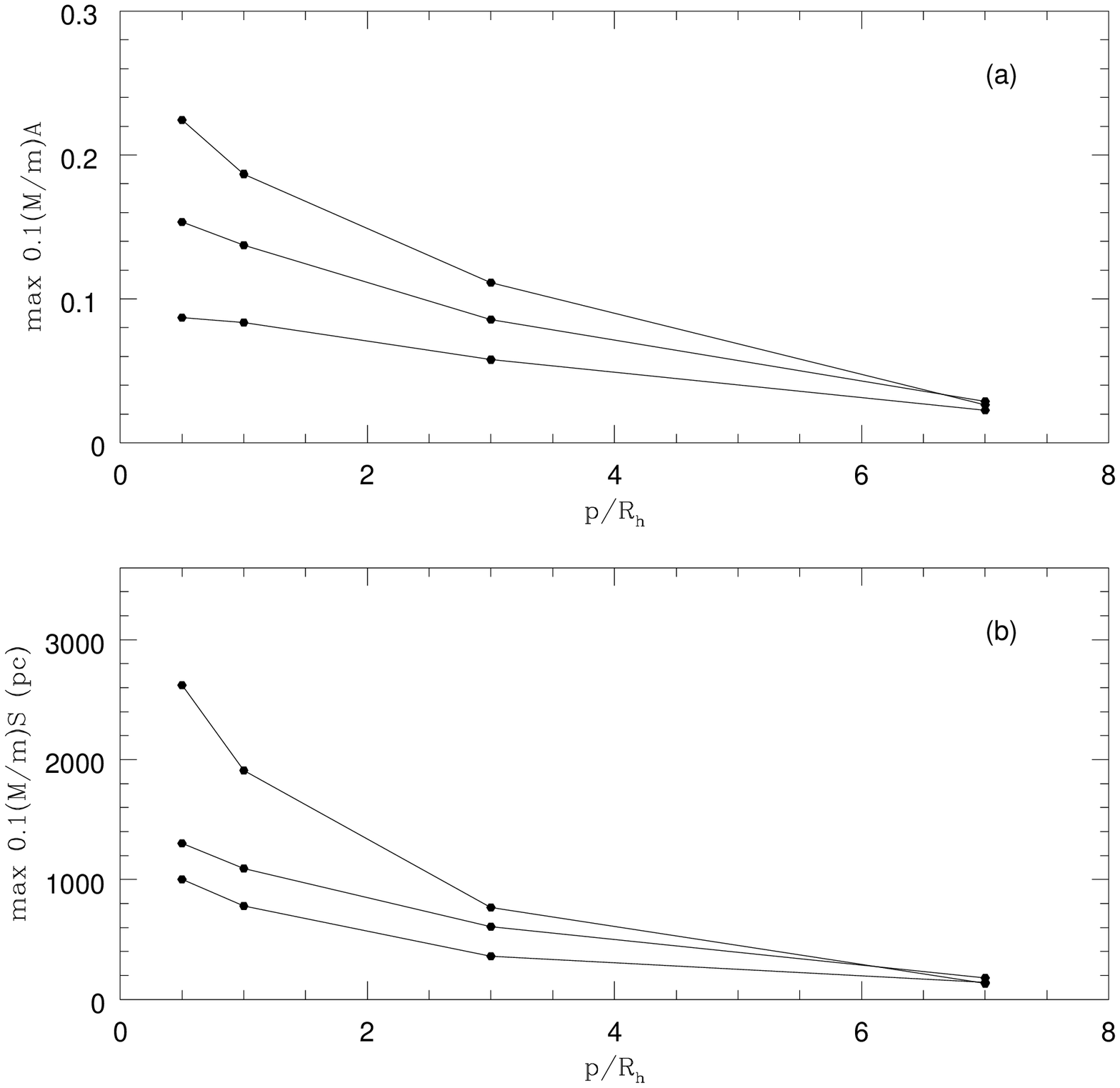}
\caption{Maximum value taken by $A$ (panel a), $S$ (panel b) as a
function of the perturber pericenter of the orbit of the perturber. The three
curves in each panel corresponds (from the upper to the lower curve)
to $V=200$ km/s, $V=500$ km/s and $1000$ km/s. }
\end{figure*}
We have fit  the scaling of $\max 0.1(M/m)A$ and $\max
0.1(M/m)S$  with the relative velocity of the encounter for
different values of $p$ with a power-law $K/(V (\hbox{ km
s}^{-1})/200)^{\alpha}$ and 
we report the values obtained for the cases investigated in Table 5.
These fits are not valid in the limit of adiabatic encounters where the
strength of the perturbation  and of the diagnostic parameters 
tend to a constant value as described previously (see \S 3.1).
\subsection{Radial dependence of the asymmetry parameter $A$}
Fly-by interactions described here produce structure on well defined
scales (cf. Fig. 16), however, the sum in equation
(\ref{asym}) washes out the signature of  our 
predicted response by averaging over regions with little
distortion. To improve sensitivity, we propose an {\it incomplete}
form of the asymmetry parameter $A(r)$ which restricts the sum in
equation (\ref{asym}) to pixels enclosed by a circle of radius $r$
 from the position of the peak of density $R_p$. 
\begin{figure*}
\plotfiddle{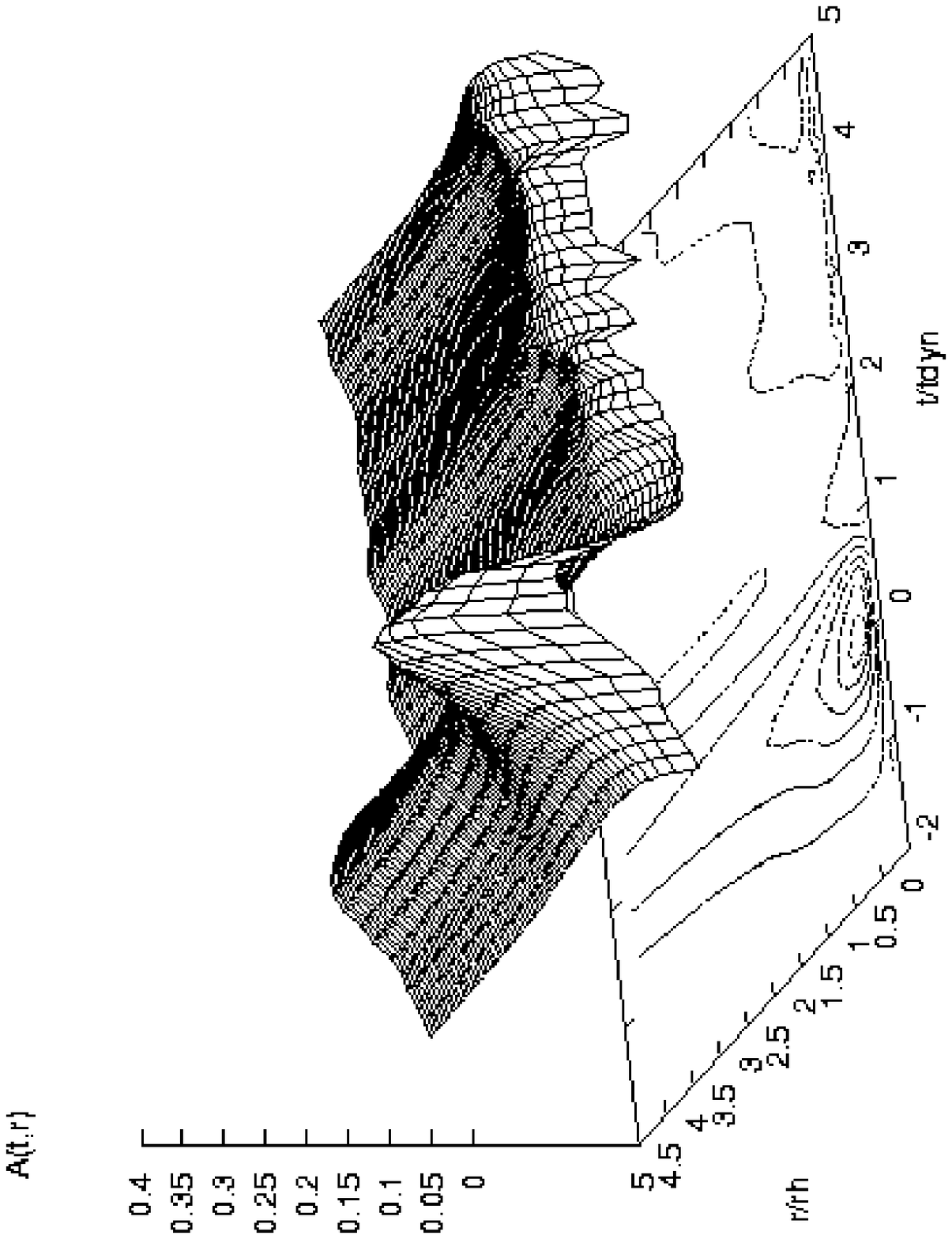}{10cm}{270}{60}{60}{-240}{320}
\caption{Plot of the asymmetry parameter, $A$, as a function of time
and radius of the area (centered at the position of the peak of
density) considered for its calculation for the fly-by with $V=500$
km/s and $p/R_h=0.5$.}
\end{figure*}
\begin{figure*}
\plotone{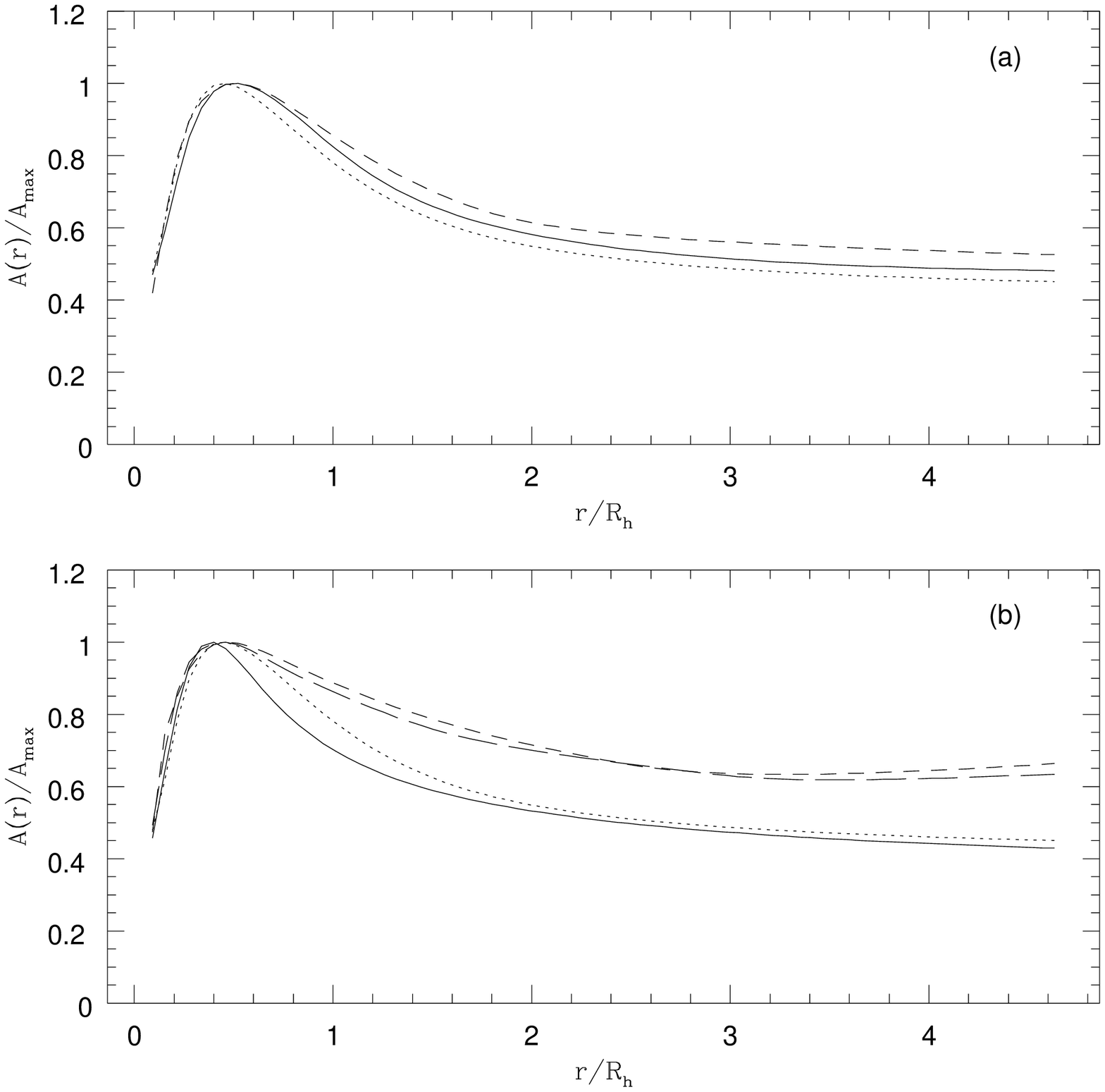}
\caption{(a) Radial profile of $A$ (normalized to its maximum value)
for fly-bys with $p/R_h=1.0$ and $V=200$ km/s (solid line), $V=500$
km/s (dotted line),
$V=1000$ km/s (dashed line); each curve correspond to the radial profile of
$A$  when the total $A$ is maximum. (b) Same as (a)
but for fly-by with $V=500$ km/s and $p/R_h=0.5$ (solid line), $p/R_h=1.0$
(dotted line), $p/R_h=3.0$ (dashed line), $p/R_h=7.0$ (long dashed line).}
\end{figure*}
Figure 21 shows the plot of $A(t,r)$ (calculated for $m/M=0.1$) for
the case with $V=500$ km/s  and $p/R_h=0.5$.
The appearance of $A(t,r)$ is qualitatively similar for the other
cases investigated.
The profile of $A$ 
at any given time has a maximum at $0.3\ltorder r/R_h \ltorder 1.0 $ 
with a subsequent decrease which is  steeper 
near the absolute maximum of $A(t,r)$ and  shallower 
far from this maximum. The value of $A$ always tends to a constant  for
large values of $r$.
In Figure 22a we compare the radial profile of $A$ for the three
fly-bys with $p/R_h=1.0$ and $V=200,~500,~1000~$ km/s. The three
curves show the run of $A(r)$ at peak and each curve is
normalized to the maximum value of $A(r)$, $A_{max}$. The radial
location of $A_{max}$ and the shape of 
$A(r)$ do not depend significantly  on the velocity of the perturber
in these half-mass units.
Figure 22b compares $A(r)/A_{max}$ for all the fly-bys with $V=500$
km/s and with the different values of $p/R_h$ investigated. The location of
the peak of $A(r)/A_{max}$ depends very slightly on $p/R_h$ while
there is a marked difference between the slope of $A(r)/A_{max}$ after the
peak for the fly-bys with small pericenters ($p/R_h<1.0$) and those with
larger values of $p/R_h$ $(p/R_h>3.0)$. 
A distinct peak in the radial profile of $A$ is present only
at some phases of the encounter and  this could be difficult to
capture in observations. 
On the other hand the decrease at small values of $r$ is a
feature common to all the 
phases of the orbit of the perturber and it should be an
observable signature.
\subsection{Summary and discussion}
In this section, we have studied the response of a high-concentration
spherical stellar system to the perturbation induced by a fly-by. 
We represent the outer profiles of cluster ellipticals by a
high-concentration King model; we choose a King profile rather than a
core-free model for numerical convenience.
A deformation in the structure of the primary galaxy is directly
observable and we explored  quantitative
measures of the asymmetry produced during the fly-by encounter. In
the end, we  focussed our attention on the 
asymmetry parameter $A$, introduced by Abraham et 
al. (1996a, 1996b) and measured for a number of local galaxies and in
galaxies in the MDS and in the HDF fields, and on the shift $S$, between the
position of the peak of the density of the 
system and its center of mass. We present the time evolution, the
scaling with the perturber orbit
and the radial dependence of the asymmetry to  facilitate estimates for a
wide variety of environments.  

The values of $A$ caused by fly-bys range from
$A\approx 0$ to $A \approx 0.2$ and this same range  of values of $A$
is found in observations of numerous local galaxies and distant
galaxies in the MDS and in the HDF fields (Abraham et al. 1996a, 1996b). 
Extreme values of $A$  observed for some galaxies in 
the HDF field ($A\gtorder0.2$) are probably caused by mergers.
The quantities $A$ and $S$ reach their maximum values when the perturber
is close to the pericenter of its orbit. As evident in
Figure 18, the distortions may be long-lived;
in some cases the perturber can be quite distant from
the primary system and therefore we can expect some non-negligible values of
$A$ and $S$ even without any close companion in the vicinity of the primary.

Our results show that the values of $A$ and $S$  will be determined by
both the distribution of relative encounter velocities  and the density of
the environment. In short, richer groups and clusters with lower
velocity dispersions will lead to  high  values of $A$ and $S$.
Figure 18 shows that disturbances persist until $\Theta \gtorder 1.4$
or roughly $\tau \sim 5 \times 10^8$ years after the pericenter passage.
In order to provide a quantitative indication of the
dependence of the response on the environmental conditions, we have
calculated the probability that a galaxy has an encounter in the time
interval $\tau$
such that $\max A \geq 0.1$ assuming that the perturber
has a mass $m=0.1M$ (or $m=0.38M$ if the projected density is used for
the calculation of $A$). 
For our fiducial system $\tau=5 \times 10^8$ yrs and
$m=10^{11}M_{\odot}$, 
the probability $P$ of having such an encounter is given by
\be
P=n\pi p^2(V)V\tau
\ee
where $n$ is the galaxy number density of the environment and $p(V)$ is the
pericenter distance for which $\max A=0.1$ during an encounter 
with velocity $V$. The function $p(V)$ has been
determined using the parametrization reported in Table 5 
for the values of $V$ available from our numerical investigation and
then fitting these points with a function of the form $K/V^{\beta}$.
We obtain
\be
P(n,V)={10^{-3}n(\hbox{Mpc}^{-3}) \over (V(\hbox{km
s}^{-1})/200)^{1.44}} 
\ee
and describe this function in Figure 23. This quantifies our finding
that 
galaxies in environments characterized by high
density and low-velocity dispersion, like compact groups, 
will be more likely to show signs of past or recent interactions in
agreement with the results of a number of investigations
(see e.g. Zepf \& Whitmore 1993, Mendes de Oliveira \& Hickson 1994).
These observational studies argue that the fraction of distorted
morphologies in compact 
groups is larger than the fraction observed in clusters (which 
have similar number densities but higher velocity dispersions) and in
sample of field galaxies (which are likely to suffer low-velocity
encounters but in a low-density environment).
Assuming that young galaxies were located in dense
environments and characterized by lower, previrialized velocity
dispersions, our conclusions are 
consistent with the results of Abraham et al. (1996b) who have  shown that the 
distribution of $A$ for galaxies in the HDF is skewed toward high
values  compared to that for galaxies in the MDS which, in turn, are more
asymmetric than local field galaxies. 

\begin{figure*}
\plotone{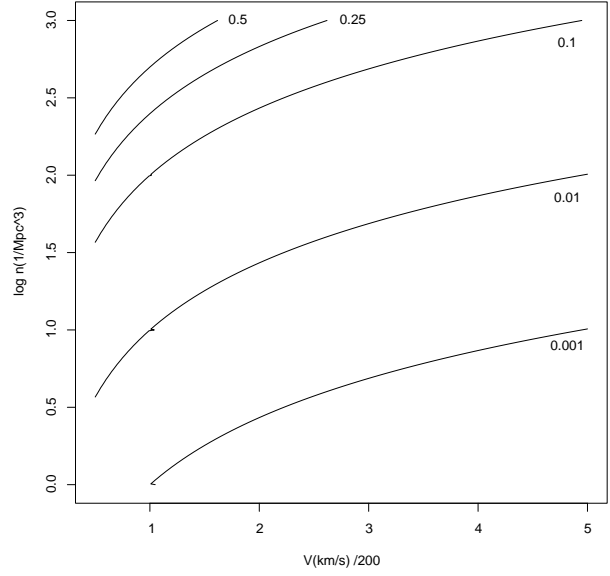}
\caption{Loci of constant probability $P(n,V)$ for an encounter in the past
$t=10t_{dyn}$  such that $\max A \geq 0.1$. The vertical and
horizontal axes describe the number density of the environment and
relative velocity of encounters.}
\end{figure*}

An additional useful diagnostic is the ratio of the
perturber  to the primary masses necessary to produce  a given value
of $\max A$. 
This is easily obtained from the fits discussed in \S 4.3 and is given by 
\be
\left({m\over M}\right)_{\max A}={0.1 \over K_A} \max A \left({V
(\hbox{km/s})\over 200}\right)^{\alpha_A},
\ee
where the values of $K_A$ and $\alpha_A$ depend on $p/R_h$ (cf. Table 5).
The four frames in Figure 24 show the contour plots of $(m/M)_{\max A}$ in the
plane $\max A-V$ (as we discussed at the beginning of this section, if
the projected density is used for the calculation of $A$ the values of
$m/M$ are larger than those reported in Fig. 24 by a
factor 3.8). Any environment
producing  low-velocity, close encounters 
can lead to significant asymmetries
($\max A \gtorder 0.1$) in the primary system even by low-mass
perturbers ($m/M \simeq 0.02-0.1$). 
For example, consider a system with a mass and a radius equal to
those of  our fiducial system ($M=10^{12}~M_{\odot}$, $R=120$ kpc). An
encounter with a perturber with $m/M=0.05$  
 with relative velocity $V=200$ km/s leads to $\max A=0.1$
if $p/R_h=1.0$ ($p\simeq 14$ kpc for our fiducial system).
Low-mass interlopers with pericenters close to the primary center 
could lead to significant asymmetries in cluster ellipticals as well. 

\begin{figure*}
\plotone{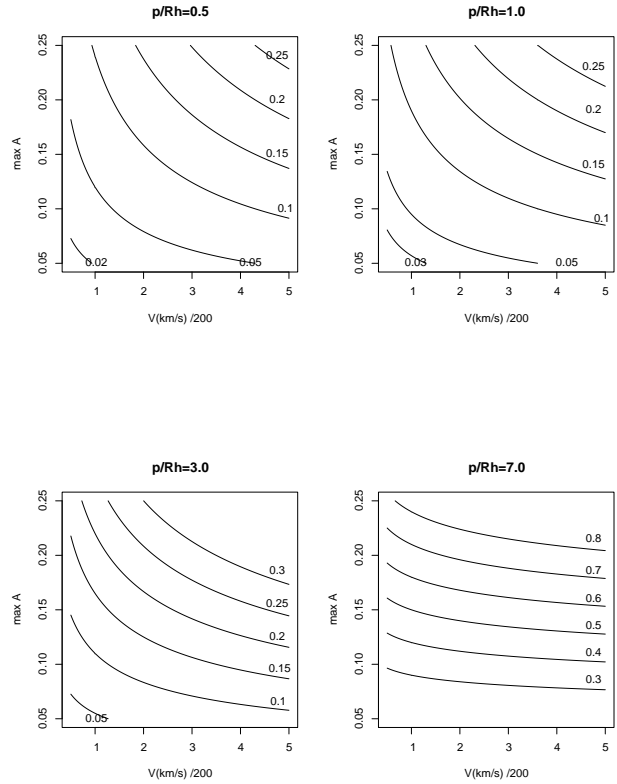}
\caption{Loci of constant relative perturber mass
necessary to have a maximum value of the asymmetry parameter equal to
$\max A$ during an encounter with relative velocity $V$. 
Pericenter distances label  each frame.}
\end{figure*}

Fly-by encounters yield the following signature in our
 radially-dependent asymmetry parameter, $A(r)$:
1) a decrease at small radii, 2) a peak around
$0.3 \ltorder r/R_h\ltorder 1.0$ 
and 3) a convergence to a constant value at large radii. 
Although we have not predicted the signature of $A(r)$ for scenarios
other than fly-bys, we believe an
observational investigation would provide a test of the mechanism
proposed here and would be a general probe of asymmetric structure. 
The computation of $A(r)$ is relatively straightforward  with the data
already available.

\section{Conclusions}
We have explored the structure and the persistence of the features in
dark halos and cluster ellipticals excited by an unbound (or {\em
  fly-by}) interaction over a range of different galactic
environments.  The main conclusions of our analysis are as follows:
\begin{enumerate}
\item Fly-by encounters which penetrate the outer regions of a
  galaxy can lead to
  asymmetric features well inside the half-mass radius. The distortion
  of a dark halo may affect its disk (e.g. Weinberg 1998b).
\item In dark halos the lowest-order damped modes yield strong features  which
  persist well after the perturber's passage and could account for
   distorted morphologies in galaxies without obvious close
  companions. Such damped modes are a natural part of the
  collisionless response but are difficult to resolve in N-body
  simulations with modest particle number ($N\ltorder 10^6$) due to noise.
\item The strength of the response depends both on the velocity of the
  perturber and on the pericenter of its orbit; low-velocity, close
  encounters produce the strongest perturbations.
\item Completely external fly-by encounters will produce a strong
  response if the perturber is more massive than the primary. For
  example such encounters can produce significant effects on spiral
  galaxies in clusters where they are likely to interact with more
  massive giant ellipticals and on dwarfs in the vicinity of normal
  galaxies.
\item A proposed generalization of the asymmetry parameter $A$
  (Abraham et al. 1996a), provides a  diagnostic of the radial
  structure of the asymmetry and a test of the mechanism explored here
  in particular.  Our predicted range of $A$ is similar to that
  observed for the majority of galaxies in the Medium Deep Survey, in
  the Hubble Deep Field and for local galaxies.  We quantify the range
  of high-density, low-velocity dispersion environments expected to
   produce significant distortions by fly-by encounters. Indeed, compact groups
  have been shown to have a significant fraction of distorted
  galaxies.
\end{enumerate}
Additional discussion is provided in \S 3.4 for dark halos and in \S
4.5 for cluster ellipticals.  

Future work will explore the properties of the response for cuspy
galaxies;  the magnitude of the resonant coupling to the inner
galaxy in core-free profiles would nicely complement the present work.
We plan to further investigate the effects of the halo distortions on
an embedded stellar and gaseous disk, by extending our current study to explore
the response of the disk and to quantify the induced deformations.
This will permit direct comparisons with the observed distortions in
the morphology of  disks as well as to study possible effects
of these distortions on their kinematics. In particular, the effects
of even small deformations in the morphologies of  disks could
significantly affect their kinematics possibly giving rise to the
observed scatter in the Tully-Fisher relation as suggested by Franx \&
de Zeeuw (1992).
\acknowledgments
We thank Eric Linder for comments on the manuscript.
This work described here was supported in part by NSF AST-9529328.

\section*{References}
Abraham R.G.,van den Bergh S., Glazebrook K., Ellis R.S., Santiago
B.X., Surma P., \& Griffiths R.E. 1996a, ApJS, 107,1\\ 
Abraham R.G., Tanvir N.R., Santiago B.X., Ellis R.S., Glazebrook K., \&
van den Bergh S. 1996b, MNRAS, 279, L47\\
Barnes, J. 1998, in Galaxies : Interactions and Induced Star Formation
: Saas-Fee Advanced Course 26 Lecture Notes 1996 Swiss Society for
Astrophysics and Astronomy)\\  
Barnes, J., \& Hernquist, L. 1992, ARA\&A, 30, 705\\
Bertin, G., \& Mark, J.W.K., 1980, A\&A, 1980, 88, 289\\
Bertin G., \& Pegoraro F., 1989, In ESA, Proceedings of
International School and Workshop on Plasma Astrophysics, Volume 1 p.329\\
Bertin G., Pegoraro F., Rubini F., \& Vesperini, E. 1994, ApJ, 434, 94\\
Brinchmann J. et al. 1998, ApJ, 499, 112\\
Butcher H, \& Oemler, A. 1978, ApJ, 219, 18\\
Butcher H, \& Oemler, A. 1984, ApJ, 285, 426\\
Couch W.J., Barger A.J. Smail I., Ellis R.S., \& Sharples R.M. 1998,
ApJ, 497, 188\\ 
Conselice C.J., \& Gallagher III, J.S. 1999, AJ, 117, 75\\
Conselice C.J., \& Bershady M. 1998, preprint, astro-ph 9812299\\
Dressler A. 1980, ApJ, 236, 351\\
Dressler A., Oemler A., Butcher H., \& Gunn J.E. 1994a, ApJ, 430, 107\\ 
Dressler A., Oemler A., Sparks W.B., \& Lucas R.A. 1994b, ApJ, 435, L23\\
Dressler et al. 1997, ApJ, 490, 577\\
Edmonds A.R. 1960, Angular Momentum in Quantum Mechanics, Princeton
University Press, Princeton, New Jersey\\
Franx M., \& de Zeeuw T. 1992, ApJ, 392, L47\\
Haynes M.P., Hogg D.E., Maddalena R.J., Roberts M.S., \& van Zee L.
1998, AJ, 115, 62\\
Hubble E., \& Humason M.L. 1931, ApJ, 74, 43\\
Kalnajs A.J. 1977, AJ, 212, 637\\
Kochanek C.S. 1996, ApJ, 457, 228\\
Kornreich D.A., Haynes M.P, \& Lovelace R.V. 1998, AJ, 116, 2154\\
Levine S.E., \& Sparke L.S. 1998, ApJ, 496, L13\\
Marleau F.R., \& Simard L. 1998, ApJ, 507, 585\\
Mendes de Oliveira C., \& Hickson P. 1994, ApJ, 427, 684\\
Miller R.H., \& Smith B.F. 1992, ApJ, 393, 508\\
Moore B., Katz N., Lake G., Dressler A., Oemler A., 1996, Nature,
379, 613\\
Moore B., Lake G., \& Katz N. 1998, ApJ, 495, 139\\
Murali, C. 1998, preprint, astro-ph 9811223\\
Murali, C., \& Tremaine S. 1998, MNRAS, 296, 749\\
Naim A., Ratnatunga U., \& Griffiths R.E., ApJ, 476, 510\\
Nelson, R.W., \& Tremaine S., 1995, MNRAS, 275, 897\\
Nelson, R.W., \& Tremaine S., 1996, in Lahav O.,Terlevich E.,
Terlevich R., eds, Proceedings of the 36th Herstmonceux Conference,
Gravitational Dynamics, Cambridge Univ. Press, Cambridge, p.73\\
Oemler A. 1974, ApJ, 194, 1\\
Palmer P.L., \& Papaloizou J. 1987, MNRAS, 224, 1043\\
Palmer P.L. 1994, Stability of Collisionless stellar systems, Kluwer
Academic Publishers, Dordrecht\\
Polyachenko V.L., \& Shukhman I. 1981, Sov. Astron., 25,533\\
Reshetnikov V., \& Combes F. 1998, A\&A, 337, 9\\
Reshetnikov V., \& Combes F. 1999, preprint, astro-ph9906063\\
Richter O., \& Sancisi R. 1994, A\&A, 290, L9\\
Rubin V., Waterman A.H., \& Kenney J.D.P., preprint, astro-ph 9904050\\
Rudnick G., \& Rix H.W. 1998, AJ, 116, 1163\\
Saha P. 1991, MNRAS, 248, 494\\
Sellwood J., \& Merritt D. 1994, ApJ, 425, 530\\
Swaters R.A., Schoenmakers R.H.M., Sancisi R., \& van Albada T.S. 1998,
preprint, astro-ph 9811424\\
Tremaine S., \& Weinberg M.D. 1984, MNRAS, 209, 729\\
van den Bergh S., Abraham R.G., Ellis R.S., Tanvir N.R., Santiago B.X., \&
Glazebrook K. 1996, AJ, 112, 359\\
Weinberg, M.D. 1989, MNRAS, 239, 549\\ 
Weinberg, M.D. 1994, ApJ, 421, 481\\
Weinberg, M.D. 1995, ApJ, 455, L31\\
Weinberg, M.D. 1998a, MNRAS, 297, 101\\
Weinberg, M.D. 1998b, MNRAS, 299, 499\\
Weinberg, M.D. 1999, AJ, 117, 629\\
Zaritsky D., \& White S.D.M. 1994, 435, 599\\
Zaritsky D., \& Rix H.W. 1997, ApJ, 477, 118\\
Zaritsky D., Smith R., Frenk C., \& White S.D.M. 1997, 478, 39\\
Zepf S.E., \& Whitmore B.C. 1993, ApJ, 418,72\\
\appendix
\section{Details of the derivation of the matrix equation}
Our results are based on an explicit solution for
the response of a spherical stellar system to the
perturbation induced by a perturber with mass $m$ on rectilinear
trajectory with pericenter $p$. Specifically, we
solve the coupled linearized collisionless 
Boltzmann  equation for the distribution function of the
primary stellar system and Poisson's equation (cf. \S 2) 

$$
{\partial f_1\over \partial t} +{\partial f_1\over \partial {\bf w}}
{\partial H_0\over \partial {\bf I}}- {\partial f_0\over \partial
{\bf I}} {\partial H_1\over \partial {\bf w}}=0, \eqno(1)
$$

$$
\nabla^2 \Phi_1=4\pi G\rho_1, \eqno(2)
$$
where the subscript $0$ denotes the equilibrium quantities and the
subscript $1$ denotes the 
first order perturbation of the distribution function $f$, and
Hamiltonian $H$. The perturbed potential, $\Phi_1$, is the sum of the 
tidal potential due to the perturber, $\Phi_p$, and the gravitational
potential of the response  to the perturbation,
$\Phi_{resp}$. 
After performing  the Laplace transform in time, the Fourier transform in
the angle variables and solving for $f_1$, equation (\ref{veq}) becomes

\be
\tilde{f}_{{\gb}1}=i{\gb} {\partial f_0\over \partial
{ \bf I}}{\tilde{\Phi}_{p,{\gb}}+\tilde{\Phi}_{resp,{\gb}}
\over s+i{\gb}\cdot{\ob}}, \label{lftr}
\ee
where we denote the Laplace transform of a quantity
$y(t)$ by $\tilde{y}(s)$ and the Fourier transform of a quantity
$y({\bf w})$ by $y_{\gb}$  with ${ \gb}=(\g_1,\g_2,\g_3)$ indicating the
vector of integer indices for 
the discrete Fourier expansion in the angle variables. The frequencies
associated to the angle variables are denoted by
${\ob}=\partial H_0/ \partial {\bf I}$.  

Following the approach adopted in Weinberg (1989),  
we will use a potential-density biorthogonal basis, $u^l_i,~d^l_i$ to
expand the 
perturbed quantities $\Phi_1$ and $\rho_1$. 
The pairs $u_i^l$, $d_i^l$ are solutions of the Poisson
equation. The basis we have adopted is obtained by 
numerical solution of the associated Sturm-Liouville problem
in which the lowest order basis functions are tailored to the
background equilibrium model following
Weinberg (1999). The solution to the coupled equation, which determines
the response of the halo to the
perturbation, then reduces to the solution of an algebraic integral
equation for  the coefficients of the expansion.
The subsections below sketch  the main steps in the derivation of the algebraic
equation for the coefficients of the 
expansion from equation (\ref{lftr}) (see e.g. Weinberg 1989 for additional 
details).
\subsection{Expansion of the perturbation potential in action-angle variables}
We begin by expanding the perturbing potential, $\Phi_p$, in terms of
spherical harmonics $Y_{lm}(\theta,\phi)$ (where $\theta$ and $\phi$ denotes
the angular coordinates inside the primary stellar system).
Up to quadrupole order, the potential of a point mass
perturber can be written as
\begin{eqnarray}
\Phi_p=Gm\left[\sqrt{{2\pi\over
3}}\left(-e^{-i\Theta}Y_{1,1}+e^{i\Theta}Y_{1,-1}\right) {r_{<}\over
r_{>}^2}+\sqrt{{\pi\over 5}}\left(-Y_{2,0}+
\sqrt{{3\over
2}}e^{-2i\Theta} Y_{2,2}+\sqrt{{3\over
2}}e^{2i\Theta} Y_{2,-2}\right){r_{<}^2\over
r_{>}^3}\right] \label{eq3}
\end{eqnarray}
where $\Theta(t)=\arctan({Vt/p})$ is the position angle of the
perturber defined as the angle between the axis from the center
of the system to the pericentric position of the
perturber and the axis  from the center to the position at time $t$. We
define the usual quantities $r_{<}=\min(r,R)$,
$r_{>}=\max(r,R)$ where $r$ and $R$ are the radial
coordinate inside the stellar system and the  
distance between the center of the system and the perturber, respectively.

Following Tremaine \& Weinberg (1984),
the radial part of the perturbing potential is expanded in terms of the
basis function 
$u^l_i$ and each resulting term is expressed in the form
$u^l_{i}(r)Y_{lm}(\theta,\phi)$ as a Fourier series in the
angle variables ${\bf w}$. We finally obtain:

\begin{eqnarray}
\Phi_p&=&Gm\sum_{\gb=-\infty}^{\infty}\delta_{\gamma_3 m}\left[\sqrt{{\pi\over
5}}\left(-V_{2\g_2\g_3}\delta_{\g_30}+\sqrt{3 \over 2}e^{-2i\Theta}
V_{2\g_2\g_3}\delta_{\g_32} +\sqrt{3 \over 2}e^{2i\Theta}
V_{2\g_2\g_3}\delta_{\g_3-2}\right)X_{2\g_2\g_3}^{\g_1}+\right. \nonumber \\
& & \left. \sqrt{\left({2\pi\over 3}\right)}\left(-
e^{-i\Theta}V_{1\g_2\g_3}\delta_{\g_31}+e^{i\Theta}V_{1\g_2\g_3}\delta_{\g_3-1}\right)X_{1\g_2\g_3}^{\g_1}\right]e^{i\gb{\bf w}}
\end{eqnarray}
where 
\be 
X_{l\g_2\g_3}^{\g_1}={1\over
2\pi}\int_{-\pi}^{\pi}dw_1u_{\g_3}^l(r)e^{-i[\g_1w_1+\g_2(w_2-\psi)]}
\ee
and 
\be
V_{l\g_2\g_3}(\beta)=r^l_{\g_2\g_3}(\beta)Y_{l\g_3}(\pi/2,0)i^{\g_3-\g_2}.
\ee
The angle $\beta$ is the inclination of the orbital plane to the
equatorial plane and $r^l_{\g_2\g_3}(\beta)$ are the components of the
rotation matrices (see e.g. Edmonds 1960).
\subsection{Derivation of matrix equation}
Similarly, we can express the perturbing and the response
potential in the form 
\be 
\Phi=\sum_{l,{\gb}}V_{l\g_2\g_3}(\beta)\sum_j
a^j_{l\g_3}(t)X_{l\g_2\g_3}^{\g_1,j}e^{i{\gb \cdot \bf w}}.
\ee
For the perturbing potential, each coefficient
$a^j_{l\g_3}(t)$ can be expressed as the product of a
 time-independent coefficient, $b^j_{l\g_3}$, and a function of time
depending only on $l$ and $\g_3$, $g_{l\g_3}(t)$. The functional form of
$g_{l\g_3}(t)$ can be easily derived for any value of $l$ and $\g_3$ using
equation (\ref{eq3}). For example, for $r<R$, $l=2$ and $\g_3=2$, one finds
$
g_{22}=\cos^3\Theta(t)e^{-2i\Theta(t)}.
$
Equation (\ref{lftr}) now takes the form
\begin{eqnarray}
\tilde{f}_{\gb 1}={i \gb \cdot {\partial f_0\over \partial
{\bf I}}\over s+i\gb \cdot {\ob}}\left[\sum_{l}V_{l\g_2\g_3}(\beta)\sum_j 
\tilde{a}^j_{l\g_3}(s)X_{l\g_2\g_3}^{\g_1,j}e^{i{\gb \cdot \bf w}} + \sum_{l}V_{l\g_2\g_3}(\beta)\sum_j 
b^j_{l\g_3}\tilde{g}_{l\g_3}(s)X_{l\g_2\g_3}^{\g_1,j}e^{i{\gb \cdot \bf w}}\right].
 \label{lftr1}
\end{eqnarray}
For self-consistency, the response density, which is obtained by
integrating equation (\ref{lftr1}) over the velocity coordinates,
$\rho_1^{resp}$,  must be
equal to the density  $\rho_1^{P}$ determined by the Poisson's equation with
$\Phi_1=\Phi_1^{resp}$. 
Imposing the condition
\be
\tilde{\rho_1}^P=\tilde{\rho_1}^{resp}, \label{cons}
\ee 
multiplying both
sides of equation(\ref{cons}) by $Y_{l\g_3}^*u^l_i$ 
and integrating over the spatial coordinates, we obtain the
following equation for the coefficients 
\be
\tilde{a}^i_{l\g_3}(s)=\sum_jM^{ij}_{l\g_3}(s)[\tilde{a}^j_{l\g_3}(s)+b^j_{l\g_3}\tilde{g}(s)]\label{alapl}.
\ee
Equivalently, this may be written as a matrix equation:
\be
{\bf \tilde{a}}_{l\g_3}(s)={\bf D}_{l\g_3}^{-1}(s){\bf M}_{l\g_3}(s)[{\bf b}_{l\g_3}\tilde{g}(s)]\label{alapl1}
\ee
where the elements of the  matrix ${\bf M}_{l\g_3}(s)$ are defined as
\be
M^{ij}_{l\g_3}(s)\equiv\sum_{\g_1,\g_2}\int \hbox{d}{\bf I}\hbox{d}{\bf w}
{i\gb \cdot {\partial f_0\over \partial {\bf I}}\over s+i\gb \cdot 
{\ob}}V^{*}_{l-\g_2-\g_3}V_{l\g_2\g_3}X_{l-\g_2-\g_3}^{\g_1,i*}
X_{l\g_2\g_3}^{\g_1,j} 
\ee 
and 
\be 
{\bf D}_{l\g_3}\equiv {\bf I}-{\bf M}_{l\g_3}.
\ee
Equations (\ref{alapl}) or (\ref{alapl1})
completely determines the response to the perturbation induced by the
perturber.
For ${\bf b}=0$ (no external perturbation), one obtains the following nonlinear
eigenvalue problem 
\be
\tilde{a}^i_{l\g_3}(s)=\sum_jM^{ij}_{l\g_3}(s)[\tilde{a}^j_{l\g_3}(s)].
\label{al}
\ee
The eigenvalues of this problem are the frequencies of the  point modes
of the stellar system: frequencies with $Re(s)>0$ correspond to
unstable growing modes while for stable damped modes $Re(s)<0$.  

\subsection{Inverse Laplace transform}
Solution of equation (\ref{alapl}) describes the evolution in frequency
space. In order to get the time evolution of the perturbation we need
to perform the inverse Laplace transform of equation (\ref{alapl}). This
yields the following
expression for the coefficients ${\bf a}_{l\g_3}(t)$
\begin{eqnarray}
{\bf a}^i_{l\g_3}(t)=\left[\sum_{\g_1,\g_2}\int \hbox{d} {\bf I}\hbox{d} {\bf
w}D^{-1}_{ik}(-i{\gb}\cdot {\ob})\mu^{kj}e^{-i {\gb}\cdot {\ob} t}\int_{-\infty}^t g_{l\g_3}(t^{'})e^{i {\gb}\cdot{\ob} t^{'}} \hbox{d} t^{'}\right]b^j_{l\g_3}, \label{tco}
\end{eqnarray}
where, to simplify the notation, we have defined 
\be
\mu^{kj}\equiv 
{i\gb \cdot {\partial f_0\over \partial \bf{I}}}
V^{*}_{l-\g_2-\g_3}V_{l\g_2\g_3}X_{l-\g_2-\g_3}^{\g_1,i*}
X_{l\g_2\g_3}^{\g_1,j}. 
\ee
Equation (\ref{tco}) does not take into account the  zeros of 
det ${\bf D}(s)$ which are the point modes of the primary system. Since we will
consider only systems known to be stable, there are  no modes with
$Re(s)>0$, while modes with $Re(s)<0$ will be damped and die away with
time. Nevertheless, as shown in Weinberg (1994), for King models 
some modes can be very weakly damped
and persist long after their excitation. When the effects of damped
modes are included, Equation (\ref{tco}) is valid for $t \rightarrow
\infty$ while for finite time it has an
additional term for each such mode. Altogether we find 
\begin{eqnarray}
{\bf a}^i_{l\g_3}(t)&=&\sum_{\g_1,\g_2}\left[\int \hbox{d} {\bf I}\hbox{d} {\bf
w}D^{-1}_{ik}(-i{\gb}\cdot {\ob})\mu^{kj}e^{-i {\gb}\cdot{\ob} t} 
\int_{-\infty}^t g_{l\g_3}(t^{'})e^{i {\gb}\cdot{\ob} t^{'}} \hbox{d} t^{'}
+\sum_d\int \hbox{d} {\bf I}\hbox{d} {\bf
w}Res(D^{-1}_{ik}(s_d)){\mu^{kj}\over s_d+i{\gb}\cdot{\ob}} e^{s_d t}\times \right. \nonumber \\
& & \left. \int_{-\infty}^t
g_{l\g_3}(t^{'})e^{-s_d t^{'}} \hbox{d} t^{'} \right]b^j_{l\g_3}. \label{dm}
\end{eqnarray}
where $s_d$ denote the frequencies of the damped modes.

Numerical evaluation of equation (\ref{dm}) requires truncating the infinite
sum over $\g_1$ (the sums over the indices $\g_2$ and $\g_3$ range
from $-l$ to $l$; see Tremaine \& Weinberg 1984). 
After checking the convergence of the solution, the
final value adopted is $\g_{1max}=6$ for the low-concentration King model
 studied in \S 3  and $\g_{1max}=10$ for the high-concentration
King model considered in \S 4.
Similarly, the sum over the radial basis must be truncated. Since the 
the adopted biorthogonal functions are
tailored to the equilibrium model $n_{max}=10$ is sufficient for the
convergence. All the integrals have been calculated by
Romberg's method with the number of points chosen so to guarantee a
maximum error of $\approx10^{-4}$.  
\end{document}